\documentclass[10pt,twocolumn]{IEEEtran}
\usepackage{graphicx,latexsym}
\newcommand{\figura}[3]
{
\begin{figure}
  \centering
 \includegraphics[width=8cm]{#1}
  \caption{#2}\label{#3}
\end{figure}
}

\title{The Problem of Localization in Networks of
Randomly Deployed Nodes: Asymptotic and Finite Analysis, and
Thresholds}
\author{Fred~Daneshgaran,~
M.~Laddomada,~and 
M.~Mondin
\thanks{Fred Daneshgaran is with ECE Dept., CSU,
Los Angeles, USA.}
\thanks{Massimiliano Laddomada and Marina Mondin
are with DELEN, Politecnico di Torino, Italy.}
}
\begin{document}
\maketitle
\begin{abstract}
Consider a two dimensional domain $S\subseteq\Re^2$ containing two
sets of nodes from two statistically independent uniform Poisson
point processes with constant densities $\rho_L$ and $\rho_{NL}$.
The first point process identifies the distribution of a set of
nodes having information about their positions, hereafter denoted
as L-nodes (Localized-nodes), while the other is used to model the
spatial distribution of nodes which need to localize themselves,
hereafter denoted as NL-nodes (Not Localized-nodes). For
simplicity, both kind of nodes are equipped with the same kind of
transceiver, and communicate over a channel affected by shadow
fading.

As a first goal, we derive the probability that a randomly
chosen NL-node over $S$ gets localized as a function of a variety
of parameters. Then, we derive the probability that the whole
network of NL-nodes over $S$ gets localized.

As with many other random graph properties, the localization
probability is a monotone graph property showing thresholds. We
derive both finite (when the number of nodes in the bounded domain
is finite and does not grow) and asymptotic thresholds for the
localization probability.

In connection with the asymptotic thresholds, we show the presence
of asymptotic thresholds on the network localization probability
in two different scenarios. The first refers to dense networks,
which arise when the domain $S$ is bounded and the densities of
the two kinds of nodes tend to grow unboundedly. The second kind
of thresholds manifest themselves when the considered domain
increases but the number of nodes grow in such a way that the
L-node density remains constant throughout the investigated
domain. In this scenario, what matters is the minimum value of the
maximum transmission range averaged over the fading process,
denoted as $d_{max}$, above which the network of NL-nodes almost
surely gets asymptotically localized.

\end{abstract}
\section{Introduction and Literature overview}
This paper deals with a network composed of two sets of nodes
randomly distributed over a two dimensional domain
$S\subseteq\Re^2$ following two statistically independent Poisson
point processes with intensities $\rho_L$ and $\rho_{NL}$. The
first process is associated with the nodes that have a-priori
knowledge about their position (these are the so called L-nodes),
while the other point process is associated with the nodes that
are trying to localize themselves (these are the so called
non-localized or NL-nodes). In particular, the paper focuses on
the connection between some system level parameters and the node
localization probability in a Poisson distributed configuration of
nodes, which are at the basis of topological network control. We
do not propose any new or modified localization method. As it will
become clear later, the primary assumptions in our analysis are:
a) nodes are Poisson distributed over a bounded circular domain
contained in $\Re^2$ and b) each node has an average typically
circular footprint representing its radio coverage. Hence, while
we focus on a particular example involving range measurements
using Received Signal Strength (RSS), the analysis can be applied
to other range measurement methods as well. Notice that Poisson
point processes are useful for modelling scenarios in which the
deployment area, the number of deployed nodes, or both, are not
\textit{a-priori} known. The Poisson model is in fact a good
approximation of a binomial random variable when the number of
deployed nodes over a bounded domain is high while the node
density is constant across the whole region of interest
\cite{papoulis}. Nevertheless, the Poisson approximation leads in
many cases of interest to a mathematically tractable problem.

This general framework can be recognized in many practical
scenarios. A possible example is a Distributed Sensor Network
(DSN), in which one may be interested in distributed power
efficient algorithms to derive localization information in a
randomly distributed collection of severely energy and computation
power limited nodes. A second example may be that of a wireless
network, in which the various network elements may communicate
between themselves (in the case of wireless networks allowing
peer-to-peer communication) or with a subset of nodes whose
positions are known (this is the case of classic cellular networks
and WLANs, whereby every node must communicate with at least one
base-station or access point). With this scenario in mind, let us
provide a brief overview of the localization methods that have
been proposed in the literature.

Given the great difference between the communication and
computation capability of the nodes, as exemplified by the DSN and
WLANs, algorithms developed for localization should be tailored to
the particular scenario at hand \cite{neal},\cite{fredrik}.

Practical localization algorithms can be classified in at least
two ways: centralized or distributed \cite{neal} and range-free or
based on ranging techniques \cite{chong}. The most common
techniques are based on measured range, whereby the location of
nodes are estimated through some standard methods such as
triangulation. Cramer-Rao Bounds (CRBs) on the variance of any
unbiased estimate based on the above ranging techniques are
readily available and provide a benchmark for assessing the
performance of any given algorithm \cite{sinan}, although we
should note that the derivation of the CRB itself relies on a
probabilistic model (often assumed to be Gaussian), that describes
the connection between the parameter to be estimated and the raw
observations.

In range-free localization, connectivity between nodes is a binary
event: either two nodes are within communication range of each
other or they are not \cite{neal2}. For simplicity, we may view
this event as obtained from hard quantization of, for instance, a
RSS random variable. If RSS is above a certain detection
threshold, the nodes can communicate, otherwise they cannot. Of
course, the nature of path loss and the terrain characteristics
influence both the coverage radius and the deviation of the
coverage zone from the ideal circular geometry. In a typical
scenario there may be multipath, Multiple Access Interference
(MAI) and Non Line Of Sight (NLOS) propagation conditions
\cite{neal}. Various range free algorithms have been proposed in
the literature including the centroid algorithm \cite{inchong1},
the DV-HOP algorithm \cite{inchong8}, the Amorphous positioning
algorithm \cite{inchong7}, APIT \cite{inchong4}, and ROCRSSI
\cite{chong}.

A review of various localization techniques proposed in the
literature may be found in \cite{Bulusu}. In \cite{Shang}, the
authors propose an approach based on connectivity information for
deriving the locations of nodes in a network. In \cite{Chen}, the
authors present some work in the field of source localization in
sensor networks.

A topic somewhat related to the problem dealt with in this paper
is network connectivity. This topic has received much attention
recently \cite{SantiTesto,Hekmat}. Given $n$ homogeneous nodes
independently and uniformly distributed over a region $S\subseteq
\Re^2$, a network is said to be connected if there exists a
communication link between every pair of nodes in $S$. Early work
on this topic can be found in
\cite{ChengRobertazzi,Philips,Piret}.

In \cite{ChengRobertazzi}, the authors investigated the
percolation of broadcast information in a multihop one-dimensional
radio network modeled by a Poisson spatial process. In
\cite{Philips,Piret}, the authors investigated the connectivity of
two and one dimensional networks respectively, as a function of
the transmission range of the nodes involved in the network.

The seminal work \cite{GuptaKumar2} by Gupta and Kumar
demonstrated that a network constituted by $n$ i.i.d. randomly
distributed sensors over a disk of area $S$, is asymptotically
(i.e., for $n\rightarrow\infty$) almost surely connected if the
transmission range between nodes is chosen as
$$r(n)=\sqrt{S\cdot(\log (n)+\gamma (n))/(\pi n)}$$ provided that
$\gamma (n)\rightarrow\infty$ as $n\rightarrow\infty$. A more
careful look at the asymptotic expression for $r(n)$ above would
reveal a resemblance to a known result on random graph theory
\cite{Bollobas} which states that given a set of $n$ nodes, the
random graph formed by adding an edge between any couple of nodes
with probability $p(n)$ will become connected almost surely if
$$p(n)=(\log(n)+\gamma(n))/n$$ as $n\rightarrow\infty$, provided
that $\gamma (n)\rightarrow\infty$ as $n\rightarrow\infty$.

In \cite{XueKumar} Xue and Kumar demonstrated that in a random
network of $n$ homogeneous nodes, the number of neighbors of a
randomly chosen node required for the network to be asymptotically
connected is $\Theta (\log (n))$ as $n\rightarrow\infty$. Such
results have been extended to 3-dimensional networks in
\cite{GuptaKumar}. Other works focusing on the connectivity of
random networks over bounded domains may be found in
\cite{SantiBlough}-\cite{Bettstetter}. Finally, paper
\cite{BettstetterHartmann} studies the connectivity of multihop
radio networks in log-normal shadow fading environment by looking
at the probability that a randomly chosen node is asymptotically
isolated.

The rest of the paper is organized as follows. In
Section~\ref{Section_Problem_Formulation}, we formulate the
problem at hand, present the basic assumptions for the derivations
that follow, and briefly recall the mathematical notation needed
in connection with the evaluation of the asymptotic thresholds.
Section~\ref{Section_Random_Graph_Models} recalls the mathematical
models adopted for the characterization of the transmission
channel between the two kind of nodes. The localization
probabilities are derived in
Section~\ref{Section_Localization_Probability} for a variety of
transmission parameters. Section~\ref{Section_Finite_Analysis}
investigates the presence of finite thresholds above which the
derived localization probabilities manifest large variations. This
analysis is then extended in
Section~\ref{Section_Asymptotic_Behaviour}, taking into account
the behavior of the localization probabilities for unboundedly
increasing values of the number of deployed nodes. Finally,
Section~\ref{Section_Conclusions} is devoted to conclusions.
%
%
\section{Problem Formulation and Assumptions}
\label{Section_Problem_Formulation}
\figura{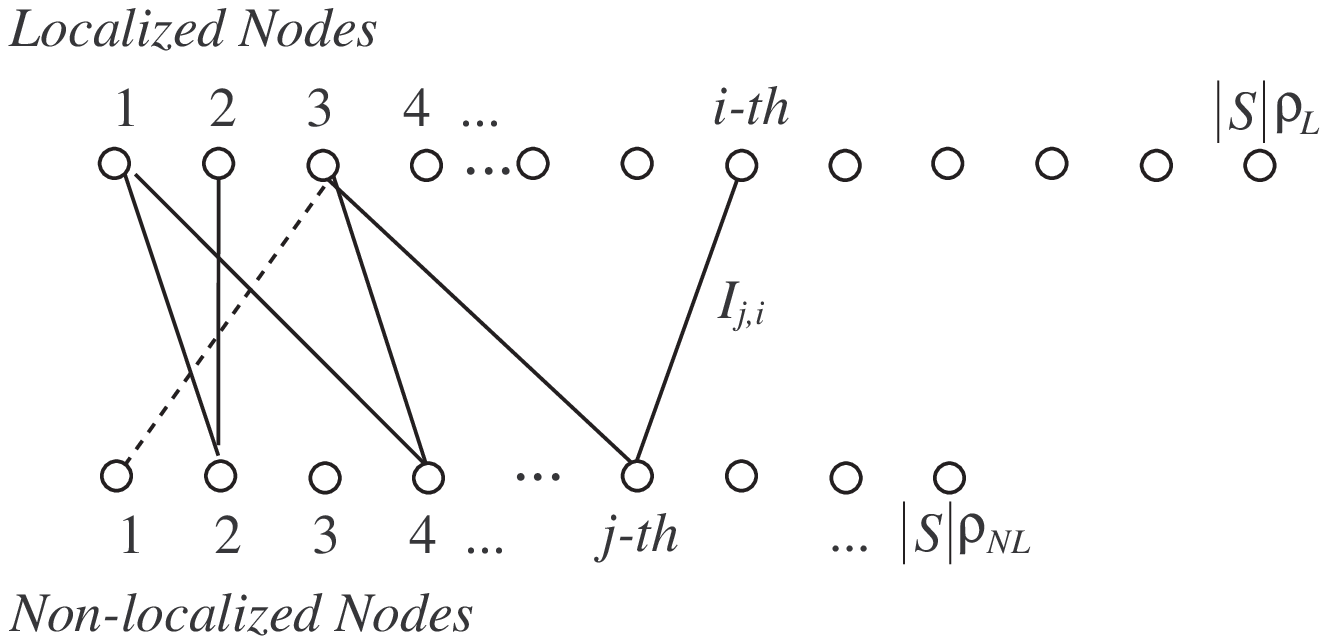}{Pictorial representation of a
bipartite network with an average number $|S|\rho_L$
of L-nodes and $|S|\rho_{NL}$ of NL-nodes over a bounded domain $S$ with
size $|S|=\pi R^2$.}{bipartite_graph}
Consider a circular domain $S\subseteq \Re^2$ of radius $R$ and
area $|S|=\pi R^2$ where sensors are deployed following two
statistically independent two dimensional Poisson point processes
with uniform densities $\rho_L$ and $\rho_{NL}$, respectively. For
simplicity, both L and NL-nodes are assumed to employ the same
kind of receiver and communicate in a scenario whereby the
transmission channel is affected by shadow fading with variance
$\sigma^2_s$. Two nodes can communicate if the received power is
above a prespecified threshold $P_{w,th}$, which is a network
parameter with respect to which the results are derived.

L-nodes have localization information relative to some coordinate
frame. Notice that how this localization is established is
irrelevant to our problem formulation.

On the other hand, NL-nodes need to localize themselves. Since we
have two kinds of nodes, the connection model between them can be
specified as a bipartite random network, denoted by
$G_{L,NL}(\rho_L,\rho_{NL})$. A pictorial representation of a
bipartite graph is shown in Fig.~\ref{bipartite_graph}, whereby an
edge between the $j$-th NL-node and the $i$-th L-node is used to
identify a communication link between the underlined nodes. Owing
to the constant densities $\rho_L$ and $\rho_{NL}$, the average
number of L and NL-nodes over $S$ is, respectively, $\rho_L\cdot
|S|$ and $\rho_{NL}\cdot |S|$.

The localization problem is two dimensional and three distance
measurements relative to nodes with known positions are sufficient
to solve for the $(X,Y)$ coordinates of the NL-node unambiguously.
%
%
\subsection{Notations}
Throughout the paper we assume the following notations
\cite{Widder}.
\begin{itemize}
    \item $x(n)=O(y(n))$ if there exists a suitable constant $c$
    such that $x(n)\le c y(n)$ for any $n\ge n_o$. Notation
    $x(n)=O(1)$ is used to signify that $x(n)$ is a bounded
    sequence.
    \item $x(n)=o(y(n))$ if $$\lim_{n\rightarrow\infty}\frac{x(n)}{y(n)}=0$$
    \item $x(n)\sim y(n)$, i.e., $x$ and $y$ are asymptotically equivalent, if and only if
    $$\lim_{n\rightarrow\infty}\frac{x(n)}{y(n)}=1$$ It is a
    matter of fact that the previous condition can also be
    represented as follows: $$x(n)=y(n)+o(y(n))=y(n)(1+o(1))$$

\item An event $E_L$ which depends on the integer-valued variable
$N$ is said to be asymptotically almost sure (a.a.s), or to occur
with high probability (w.h.p.), if
$$\lim_{N\rightarrow\infty}P(E_L)=1$$

\end{itemize}
%
%
%
\section{Random Graph Models for Wireless Networks of Randomly distributed Nodes}
\label{Section_Random_Graph_Models}
Connections between the two classes of nodes depend on the
considered channel model. Basically, three basic models have been
extensively adopted in the literature for wireless networks
analysis, namely random geometric graphs~\cite{PenroseTesto},
path-loss channel model~\cite{rappaport}, and path-loss geometric
model with shadowing~\cite{rappaport,Hekmat,BettstetterHartmann}.
\subsection{Random Geometric Graphs}
A random geometric graph suitable for the problem at hand, is
defined as follows. Let $\left(x^{NL}_{j,1},x^{NL}_{j,2}\right)$
identify the geometric position of the $j$-th NL-node,
$X^{NL}_{j}$, with $j=1,\ldots,\rho_{NL}|S|$, and let $D=\|\cdot
\|$ be some suitable norm\footnote{A thoroughly employed norm is
the Euclidean norm.} on $\Re^2$. In a random geometric graph,
$X^{NL}_{j}$ is connected to a L-node $X^L_{i}$ with
$i=1,\ldots,\rho_L|S|$ over the domain $S$ by an undirected edge
if $D=\|X^{NL}_{j}-X^{L}_{i}\|\le r$, whereby $r$ is some positive
predefined parameter.

This is a reasonable assumption in practice. In fact, usually
receivers have strict signal-to-noise (SNR) requirements such that
if the SNR is above a predefined threshold, i.e., if the distance
between the nodes is below a given value, then reliable
communication between the nodes is possible; otherwise, no
communication is allowed.
\subsection{Path-loss Geometric Random Graph, Without Shadowing}
A somewhat better model accounting for practical communication
receivers is the so-called path-loss geometric random graph.

Let us assume that the $j$-th NL-node can communicate with the
$i$-th L-node if the power received by the $i$-th L-node is
greater or equal to a certain threshold $P_{w,th}$. The coverage
area of the $j$-th NL-node comprises the L-nodes where the
received power from NL-node $j$ is greater than or equal to
$P_{w,th}$. A NL-node can only communicate directly with L-nodes
that fall inside its coverage area. With this setup, we can model
the presence of a communication link between the $j$-th NL-node
and the $i$-th L-node with a random variable $I_{j,i}$ as shown in
Fig.~\ref{bipartite_graph}. $I_{j,i}$ is a discrete random
variable assuming two possible values with probabilities $P_{ji}$
and $1-P_{ji}$, i.e.
\begin{equation}\label{randomvariableI}
I_{j,i}=\left\{
\begin{array}{ll}
    1,       & P_{ji} \\
    0,       & 1 -P_{ji}
\end{array}\right.
\end{equation}
Based on the observations above, the probability
$P_{ji}=P(I_{j,i}=1)$ is equal to the probability that the power
received by the $i$-th L-node is greater or equal to the power
threshold $P_{w,th}$.

Let us consider the power $P(d_{j,i})$ received by the $i$-th
L-node at a distance $d_{j,i}$ from the $j$-th
NL-node~\cite{rappaport}:
\[
P(d_{j,i})=\frac{P_tG_tG_r \lambda^2}{(4\pi)^2d^{n_p}_{j,i}}
\]
whereby, $P_t$ is the transmitted power, $G_t$ is the transmitter
antenna gain, $G_r$ is the receiver antenna gain, $n_p$ is the
path-loss exponent, and $\lambda=c/f$ is the wavelength.
Notice that this equation is not valid for $d_{j,i}=0$.

The path-loss in dB $PL$-[dB] can be expressed as:
\begin{equation}\label{received_power_0}
PL \textrm{[dB]}=10\log_{10}\left(\frac{P_t}{P(d_{j,i})}\right)=-10\log_{10}\left(\frac{G_tG_r
\lambda^2}{(4\pi)^2d^{n_p}_{j,i}}\right)
\end{equation}
Since this equation is not valid at $d_{j,i}=0$, usually it is
specified with respect to a reference distance $d_0$. In other
words, the received power $P(d_{j,i})$ at a distance $d_{j,i}$
from the transmitter is given with respect to a reference power
$P_o$ received at a distance $d_0$, usually assumed equal to 1
meter \cite{rappaport}. Such a value may be measured in a
reference radio environment by averaging the received power at a
given distance close to the transmitter. Doing so, the equation
specifying the received power $P(d_{j,i})$ is then expressed with
respect to $P_o$:
\begin{equation}\label{received_power_1}
P(d_{j,i})=P_o\cdot
\left(\frac{d_0}{d_{j,i}}\right)^{n_p}=P_o\cdot
\left(\frac{d_{j,i}}{d_0}\right)^{-n_p},~\forall d_{j,i}\ge d_0
\end{equation}
whereby $P_o$ is the signal power at a reference distance $d_o$
normalized to one for simplicity, and $n_p$ is the path loss
exponent.
%
%
In a similar fashion, if we consider the receiver threshold power
$P_{w,th}$, and define $d_{max}$ as the distance between the
transmitter and the receiver at which the received power
$P(d_{j,i})$ equals $P_{w,th}$, we can write:
%
%
%
%
\begin{equation}\label{received_power_2}
P(d_{j,i})=P_{w,th}\cdot
\left(\frac{d_{max}}{d_{j,i}}\right)^{n_p}=P_{w,th}\cdot
\left(\frac{d_{j,i}}{d_{max}}\right)^{-n_p}
\end{equation}
With this setup, the probability $P_{ji}=P(I_{j,i}=1)$ of a link
connection between a NL-node and a L-node can be evaluated as:
\begin{equation}\label{pji}
P_{ji}=\left\{
\begin{array}{ll}
    1,       & 0<d_{j,i}\le d_{max}\le R \\
    0,       & d_{max}<d_{j,i}\le R
\end{array}
\right.
\end{equation}
whereby $R$ is the radius of the area on which the network is
established. Notice that any distance must be smaller than $R$,
and that in this model the radio coverage of any node is a perfect
circular area with radius $d_{max}$. Any L-node falling in a
circle of radius $d_{max}$ from the NL-node is assumed to
communicate with the reference NL-node. In this respect, $d_{max}$
is the coverage radius of any node, and takes on the same meaning
as $r$ in the geometric random graph model described in the
previous section. The difference is that here $d_{max}$ is related
to typical transmission conditions, while $r$ in the previous
section is only interpreted as a geometric parameter.

The only parameter of interest in this model is the maximum
distance $d_{max}$. Simulation results can be given with respect
to the normalized distance $\frac{d_{max}}{R}$ in order to
highlight the dependence of the results from the ratio between the
coverage radius of any node and of the overall deployment
area.
\subsection{Wireless Channel Model: Path-loss Geometric Random Graph with Shadowing}
Practical measurements of the signal power level received at a
certain distance from a transmitter often indicate that the
path-loss in~(\ref{received_power_0}) follows a log-normal
distribution~\cite{rappaport}. From~(\ref{received_power_2}), one
easily evaluates:
\[
10\log_{10}\left(\frac{P(d_{j,i})}{P_{w,th}}\right)=10\log_{10}\left[
\left(\frac{d_{j,i}}{d_{max}}\right)^{-n_p}\right]
\]
Let us consider the normalized variables $\overline{P}(d_{j,i})$
and $\overline{d}_{j,i}$, defined as
\[
\begin{array}{l}
  \overline{P}(d_{j,i})=  \frac{P(d_{j,i})}{P_{w,th}}      \\
\overline{d}_{j,i}=\frac{d_{j,i}}{d_{max}}
\end{array}
\]
The log-normal model is formalized as:
\[
10\log_{10}\left(\overline{P}(d_{j,i})\right)=10\log_{10}
\left[(\overline{d}_{j,i})^{-n_p}\right]+X_s
\]
whereby, $X_s$ is a Gaussian-distributed shadowing random variable, i.e, $X_s\sim
N(\mu_s,\sigma_s^2)$ with $\mu_s=0$. With this setup, the
probability that a NL-node and a L-node establish a wireless
connection is:
\[
P\left(10\log_{10}\left(\overline{P}(d_{j,i})\right)>0\right)
\]
Notice that the underlying model becomes a path-loss geometric
random graph without shadowing upon setting $\sigma_s=0$.

By considering
$\overline{P}(d_{j,i})_{dB}=10\log_{10}\left(\overline{P}(d_{j,i})\right)$
and $\mu_d=10\log_{10} \left[(\overline{d}_{j,i})^{-n_p}\right]$,
it easily follows that:
\[
P\left(\overline{P}(d_{j,i})_{dB}>0\right)=P\left(X_s>-\mu_d\right)
\]
The latter equation corresponds to:
\[
\frac{1}{\sqrt{2\pi}\sigma_s}\int_{-\mu_d}^{+\infty}e^{-\frac{y^2}{2\sigma_s^2}}dy=
\frac{1}{2}\left[1-\textrm{erf}\left(\frac{-\mu_d}{\sqrt{2}\sigma_s}\right)\right]
\]
Upon setting $\alpha=\frac{10}{\sqrt{2}\ln (10)}$ and
$\eta=\frac{\sigma_s}{n_p}$, the previous equation can be
rewritten as follows:
\begin{equation}\label{probab_shadowing}
P\left(\overline{P}(d_{j,i})_{dB}>0\right)=\frac{1}{2}\left[1-\textrm{erf}\left(\frac{\alpha}{\eta}\ln(\overline{d}_{j,i})
\right)\right]
\end{equation}
This is the probability of establishing a wireless link between a
NL-node and a L-node given that their relative distance is
$d_{j,i}$.

Let us focus on the bipartite graph of Fig.~\ref{bipartite_graph},
and assume that the $j$-th NL-node can communicate with the $i$-th
L-node if the power received by the $i$-th L-node is greater than
or equal to a certain threshold $P_{w,th}$. The coverage area of
the $j$-th NL-node comprises the L-nodes where the power received
from the $j$-th NL-node is greater than or equal to $P_{w,th}$. A
NL-node can only communicate directly with L-nodes that fall
inside its coverage area. However, with respect to the model
described in the previous section, here there is a non-zero
probability of a wireless communication between nodes that are far
apart more than $d_{max}$ due to the considered shadow fading
model.

With the setup above, we have:
\begin{equation}\label{equat_d_max}
d_{max}=10^{\frac{\beta_{th}}{10\cdot n_p}}
\end{equation}
whereby,
\begin{equation}\label{equat_beta_thr}
\beta_{th}=10\log_{10}\left(\frac{P_t}{P_{w,th}}\right)
\end{equation}
With this setup, we can model the presence of a communication link
between the $j$-th NL-node and the $i$-th L-node with a random
variable $I_{j,i}$ as shown in Fig.~\ref{bipartite_graph}. The
random variable $I_{j,i}$ is a discrete random variable assuming
two possible values with probabilities $P_{ji}$ and $1-P_{ji}$
like in (\ref{randomvariableI}), where
%
%
\begin{equation}\label{P_ji_definitiva}
P_{ji}=P\left(\overline{P}(d_{j,i})_{dB}>0\right)
%
\end{equation}
as in~(\ref{probab_shadowing}). This is the most general model
since when $\sigma_s=0$ it becomes a path-loss geometric model.
Moreover, upon assuming $d_{max}=r$, the geometric random graph
described by Penrose~\cite{PenroseTesto} is obtained.
\section{The Localization Probability}
\label{Section_Localization_Probability}
\figura{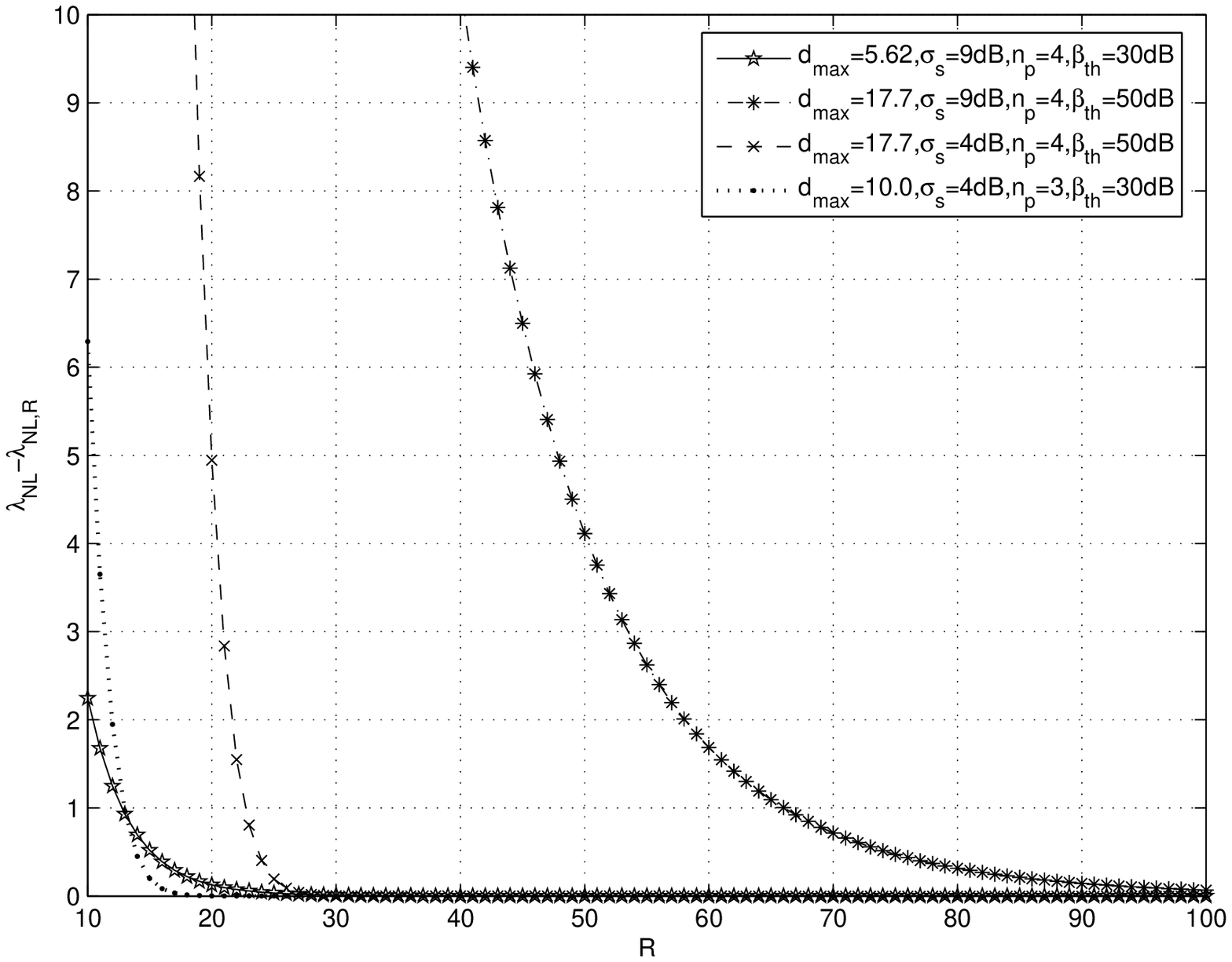}{Behavior of the
difference $\lambda_{NL}-\lambda_{NL,R}$ as a function of the
radius $R$ of the considered domain $S$. All curves are related to
$\rho_L=0.1$ nodes/m$^2$. Other transmission parameters are as
noted in the legend.}{approx_lambda_Rfinito_e_illimitato}
The aim of this section is to derive the localization probability
of the network of NL-nodes over the bounded domain $S$. The
problem is solved by first determining the localization
probability of a randomly chosen NL-node over $S$, and then upon
identifying the localization probability of the set of NL-nodes
falling within $S$ based on justifiable assumptions.

Owing to the definition of the Poisson point process describing
the NL-nodes distribution over $S$, the problem can be solved by
evaluating the expected number $\lambda_{NL,R}=E\{d^{NL}_v|R\}$ of
L-nodes seen by a NL-node within a circular area of radius $R$
centered on the NL-node. Such a random variable is denoted as
$d^{NL}_v$. Resorting to ideas from percolation theory
\cite{meester}, the expected value of neighbors within a distance
$R$ of a generic NL-node can be evaluated as follows:
\begin{equation}\label{num_neighbors}
E\{d^{NL}_v|R\}=\cdot\int_{0}^{2\pi}\int_{0}^{R}\rho_L
P\left(\overline{P}(r)_{dB}>0|r\right)rdrd\phi
\end{equation}
whereby,  $\rho_L$ is the density of the point process related to
the L-nodes, and $P\left(\overline{P}(r)_{dB}>0|r\right)$ is as
defined in (\ref{probab_shadowing}) with $r=d_{j,i}$.

The solution of (\ref{num_neighbors}), whose proof is reported in
Appendix I, is:
\begin{eqnarray}\label{num_neighbors_2}
%
\lambda_{NL,R}&=&\frac{\pi\rho_L}{2}R^2-\pi\rho_L\frac{R^2}{2}\textrm{erf}\left(\frac{\alpha}{\eta}\ln\left(\frac{R}{d_{max}}\right)
\right)\\
&+&\frac{\pi\rho_L}{2}d^2_{max}e^{\frac{\eta^2}{\alpha^2}}\left[1+\textrm{erf}\left(\frac{\alpha}{\eta}\ln\left(\frac{R}{d_{max}}\right)-\frac{\eta}{\alpha}
\right)\right]\nonumber
\end{eqnarray}
The expected number $\lambda_{NL}=E\{d^{NL}_v\}$ of L-nodes seen
by a NL-node over the entire $\Re^2$ can be evaluated as follows:
\begin{equation}\label{num_neighbors_3}
E\{d^{NL}_v\}=\lim_{R\rightarrow\infty}\int_{0}^{2\pi}\int_{0}^{R}\rho_L\cdot
P\left(\overline{P}(r)_{dB}>0|r\right)rdrd\phi
\end{equation}
The solution of (\ref{num_neighbors_3}), whose proof is given in
Appendix I, is:
\begin{equation}\label{num_neighbors_4}
\lambda_{NL}=E\{d^{NL}_v\}=\rho_L\pi
d^2_{max}e^{\frac{\eta^2}{\alpha^2}}
\end{equation}
Before proceeding further, notice that so long as $R\gg d_{max}$,
the average number of L-nodes estimated by (\ref{num_neighbors_2})
over $S\subseteq \Re^2$ coincides with the ones estimated by
(\ref{num_neighbors_4}) over the whole two dimensional domain
$\Re^2$. This is clearly depicted in Fig.
\ref{approx_lambda_Rfinito_e_illimitato} as a function of the
radius $R$ of the considered domain $S$, for a variety of
transmission parameters as noted in the legend. Actually, the less
stringent condition $R\ge 5\cdot d_{max}$ suffices to ensure
$\lambda_{NL}\approx\lambda_{NL,R}$. Owing to this observation,
when not differently specified, in what follows we will consider
the formula (\ref{num_neighbors_4}).

The next line of pursuit consists in the definition of the
localization probability of a randomly chosen NL-node within $S$.
Since L-nodes are distributed as a Poisson point process, the
number of L-nodes $d^{NL}_v$ is a Poisson random variable with
expected value $\lambda_{NL}=E\{d^{NL}_v\}$ in
(\ref{num_neighbors_4}) if $S=\Re^2$, or $\lambda_{NL,R}$ in
(\ref{num_neighbors_2}) if $S$ is a bounded domain of radius $R$
contained in $\Re^2$. The event of interest, identified by $E_L$,
is the event that a randomly chosen NL-node is within the
transmission range of at least three L-nodes. Over $\Re^2$, such a
probability can be evaluated as the probability that the random
variable $d^{NL}_v$ takes on values greater than or equal to $3$:
\begin{eqnarray}\label{loc_probability_R2}
P\left(E_L\right)&=&P\left(d^{NL}_v\ge
3\right)=\sum_{j=3}^{+\infty}\frac{E\{d^{NL}_v\}^j}{j!}e^{-E\{d^{NL}_v\}}\nonumber\\
&=&1-\sum_{j=0}^{2}\frac{E\{d^{NL}_v\}^j}{j!}e^{-E\{d^{NL}_v\}}
\end{eqnarray}
which can be rewritten as:
\figura{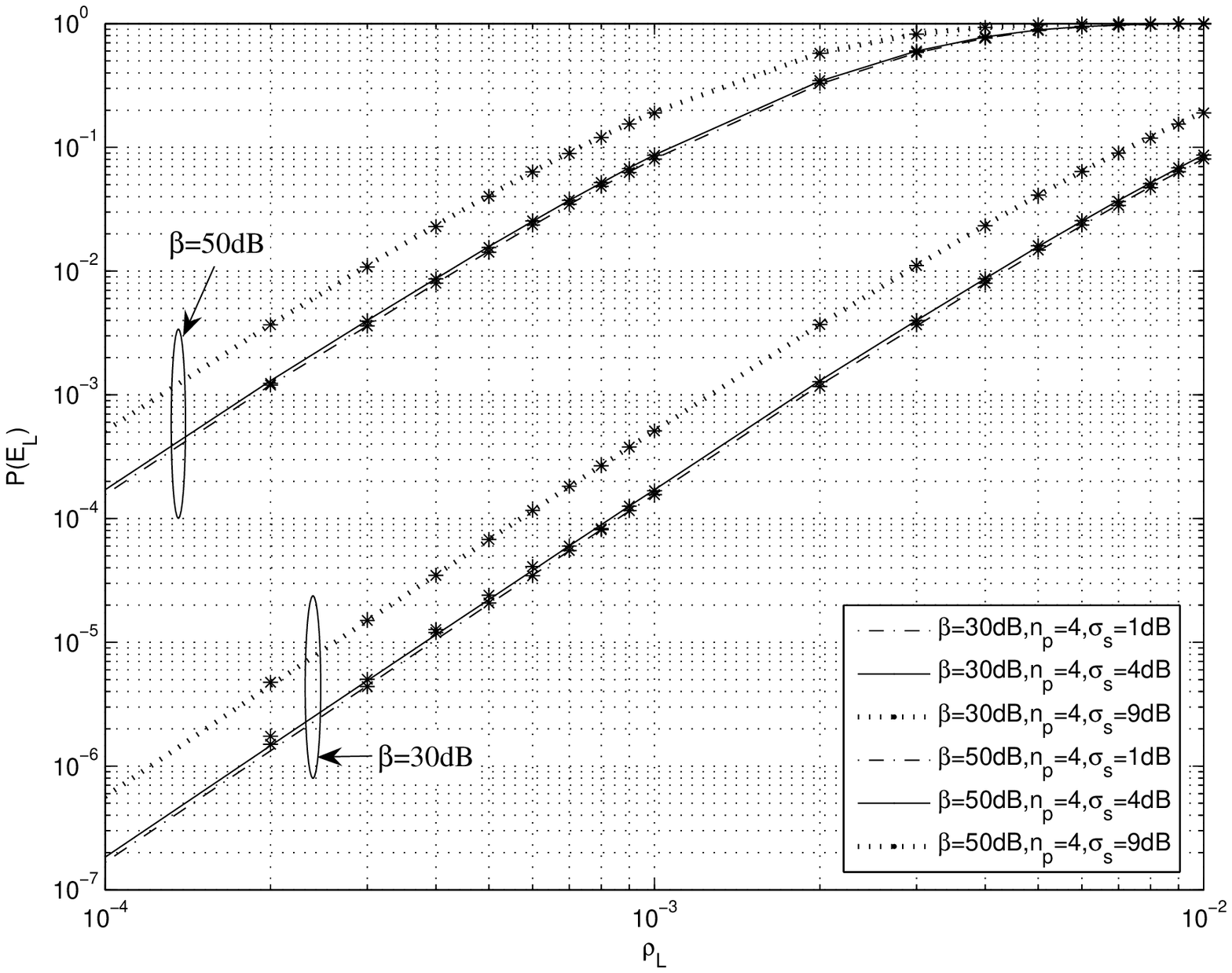}{Behavior of the localization probability
$P(E_L)$ as a function of the L-node density $\rho_L$ over
$\Re^2$. Other transmission parameters are as noted in the legend,
while $\rho_{NL}=\rho_L$. Simulated points are identified by
star-marked points over the respective theoretical curves.}{P_loc}
\[
P\left(E_L\right)=1-e^{-E\{d^{NL}_v\}}\left[1+E\{d^{NL}_v\}+\frac{E\{d^{NL}_v\}^2}{2}\right]
\]
Using (\ref{num_neighbors_4}), it is straightforward to
obtain:
\begin{equation}\label{loc_probability_R2_2}
\begin{array}{lll}
P\left(E_L\right)&=&1-e^{-\rho_L\pi
d^2_{max}e^{\frac{\eta^2}{\alpha^2}}}\left[1+\rho_L\pi
d^2_{max}e^{\frac{\eta^2}{\alpha^2}}+\right.\\
&&+\left.\frac{\rho_L^2}{2}\pi^2
d^4_{max}e^{2\frac{\eta^2}{\alpha^2}}\right]
\end{array}
\end{equation}
The behavior of $P\left(E_L\right)$ is displayed in
Fig.~\ref{P_loc} for the parameters noted in the legend.

Simulation results have been obtained as follows. We define a
square domain $C$ with size $R_d\times R_d$ and centered a
circular domain $S$ of area $\pi R^2$ in the middle of $C$. In
order to simulate the entire domain $\Re^2$, we assume $R_d\gg R$.
Furthermore, we must have $R\gg d_{max}$ in the investigated
scenario, say $R>10 d_{max}$, based on the considerations stated
above. Then, we generate two statistically independent point
processes distributed uniformly over $C$ with densities $\rho_L$
and $\rho_{NL}$, respectively. Owing to the constant density of
both point processes within $C$, the number of L-nodes falling in
$C$ is, on average, $E_C=\rho_L\cdot R_d^2$, while the average
number of L-nodes falling in $S$ is $E_R=\rho_L\cdot \pi
R^2\Rightarrow \rho_L=E_R/\pi R^2$. Upon substituting $\rho_L$ in
$E_C$ the following relation follows:
\figura{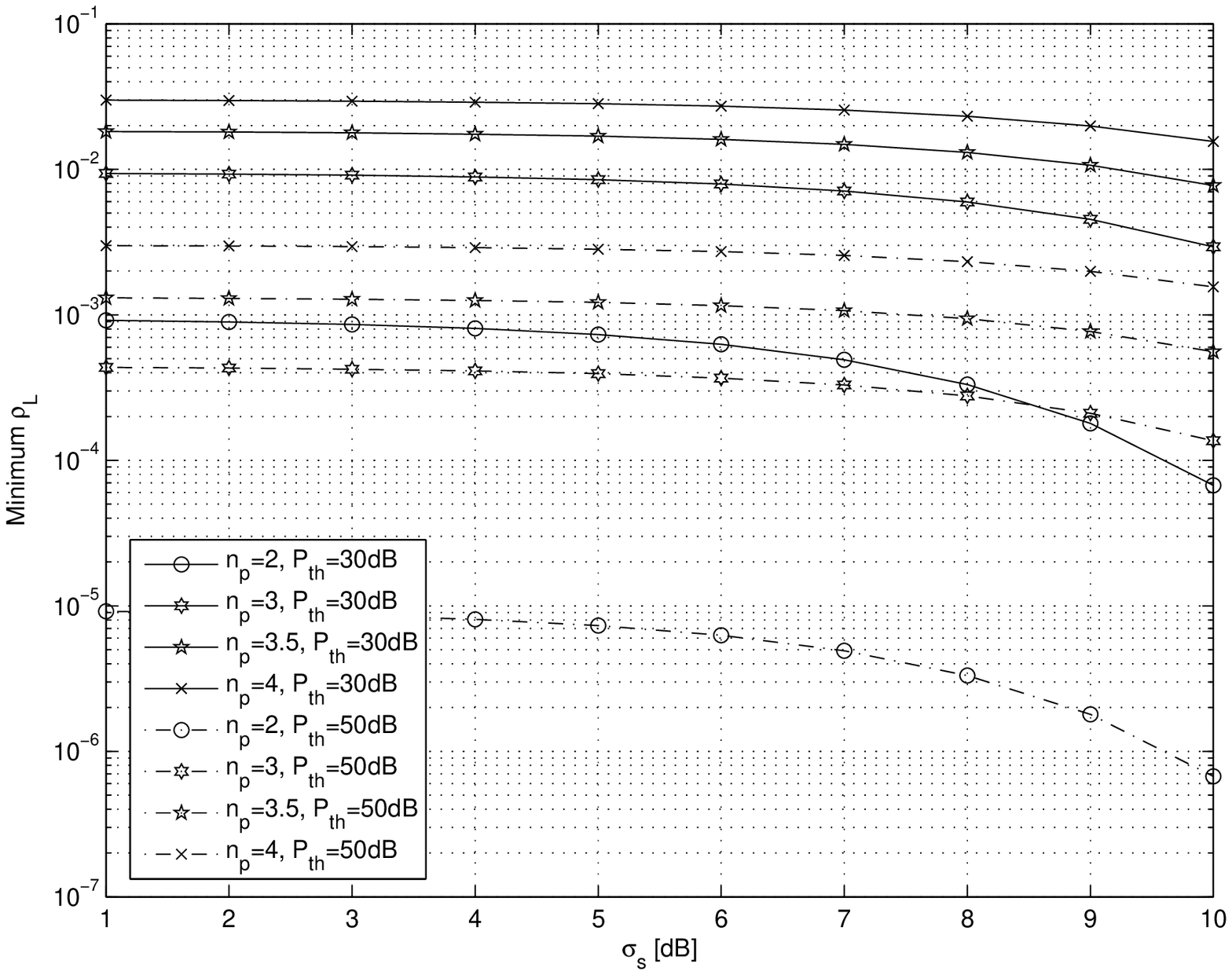}{Minimum L-node density over $\Re^2$ as a
function of $\sigma_s$ (in dB) for assuring that on the average,
each NL-node is able to establish a wireless link with at least
three neighbors under the channel conditions exemplified by the
parameters $P_{th}$ and $n_p$.}{min_rho_L}
\[
E_C=E_R\cdot \frac{R_d^2}{\pi R^2}
\]
For ensuring an appropriate number of L-nodes in $S$, say $E_R\sim
500$, $E_C$ nodes are uniformly distributed on the bigger domain
$C$. The localization probability is then evaluated by dividing
the number of localization events in the domain $S$ by the number
of randomly generated network realizations. In order to avoid
border effects, NL-nodes close to the border of the domain $S$ are
allowed to communicate with L-nodes within an annulus of radius
$d_{max}$ from the circular domain $S$.

Some observations from the results in Fig.~\ref{P_loc} are in
order. As expected, the node localization probability increases for
increasing values of the density $\rho_L$ of the L-nodes.
For fixed values of $\rho_L$, the node localization
probability increases for increasing values of the parameter
$\beta_{th}$, which in turn depends on the maximum transmission
range $d_{max}$. Moreover, note that for a given
set of transmission parameters, the localization probability
increases for increasing values of the variance of the shadow
fading $\sigma_s$.

The analysis above is the starting point for finding theoretical
conditions assuring that the localization probability is above a
certain threshold. Upon imposing $E\{d^{NL}_v\}\ge 3$, one easily
finds:
\begin{equation}\label{density_minimum}
\rho_L\ge \frac{3}{\pi
d^2_{max}}e^{-\frac{1}{\alpha^2}\frac{\sigma^2_s}{n_p^2}}
\end{equation}
which yields the minimum uniform L-node density over $\Re^2$ for
assuring that on the average each NL-node is able to establish a
wireless link with at least three neighbors under the channel
conditions exemplified by the parameters $\sigma_s$ and $n_p$.

The behavior of~(\ref{density_minimum}) as a function of the
shadowing parameter $\sigma_s$ (in dB) is displayed in
Fig.~\ref{min_rho_L} for the transmission parameters noted in the
legend. Notice that, as expected, shadowing tends to decrease the
L-node density since farther nodes can communicate over longer
distances.

The behavior of the expected number $\lambda_{NL}=E\{d^{NL}_v\}$
of L-nodes seen by a NL-node over $\Re^2$ (see
(\ref{num_neighbors_4})) is displayed in Fig.~\ref{E_d_v} as a
function of the L-node density $\rho_L$ for a variety of
transmission parameters, as summarized in the figure legend.
Star-marked points denote simulated points.
\figura{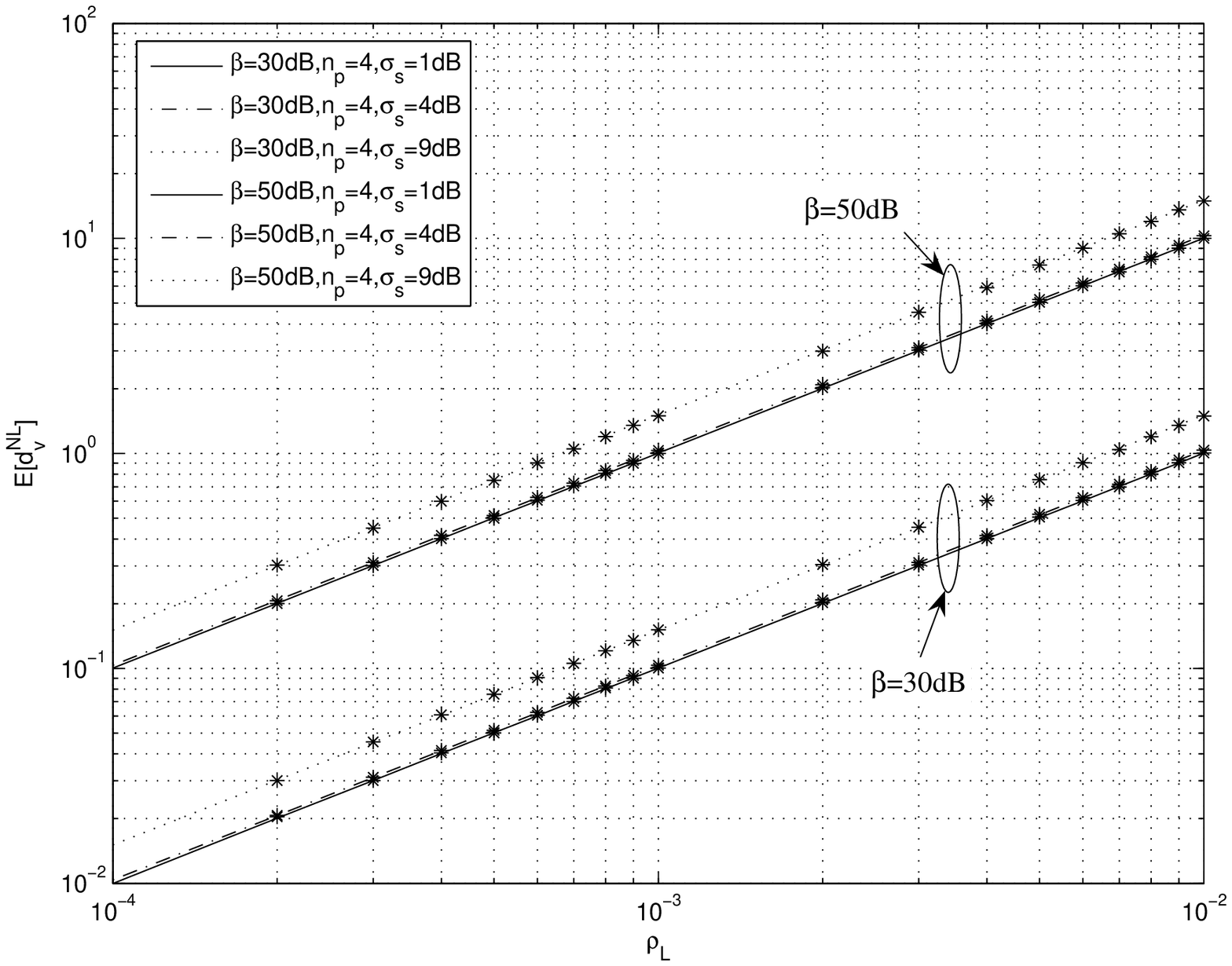}{Expected number $\lambda_{NL}=E\{d^{NL}_v\}$ of
L-nodes seen by a NL-node over $\Re^2$ (see
(\ref{num_neighbors_4})) as a function of the L-node density
$\rho_L$, for the transmission parameters noted in the legend.
Simulated points are identified by star-marked points over the
respective theoretical curves.}{E_d_v}

Next, consider the probability that the whole network of NL-nodes
falling in the bounded domain $S$ under investigation gets
localized. Such an event occurs when all the single NL-nodes
within $S$ get localized. Let $N_{NL}$ be the number of NL-nodes
falling within $S$.

Consider $P(E_L)$ in~(\ref{loc_probability_R2_2}), and define
$X(\lambda_{NL})$ as
\begin{equation}\label{temp_prob0}
\begin{array}{lll}
X(\lambda_{NL})&=&1-P(E_L)=e^{-\rho_L\pi
d^2_{max}e^{\frac{\eta^2}{\alpha^2}}}\left[1+\right.\\
&&+\rho_L\pi
d^2_{max}e^{\frac{\eta^2}{\alpha^2}}+\left.\frac{\rho_L^2}{2}\pi^2
d^4_{max}e^{2\frac{\eta^2}{\alpha^2}}\right]
\end{array}
\end{equation}
With this setup, by virtue of the independence of the NL-nodes in
$S$, the probability $P_N\left(E_L\right)$ that all the network of
NL-nodes deployed in $S$ gets localized can be expressed as:
\begin{equation}\label{loc_probability_allnetw0}
\begin{array}{lll}
P_N\left(E_L\right)&=&\left[1-X(\lambda_{NL})\right]^{N_{NL}}
\end{array}
\end{equation}
whereby, we have to interpret such a probability as conditioned on
the number of NL-nodes falling in the domain $S$. On average,
$N_{NL}=\rho_{NL}\pi R^2$ in the observation area $S$.
\section{Analysis of the Localization Probability and Thresholds, Finite Case}
\label{Section_Finite_Analysis}
\figura{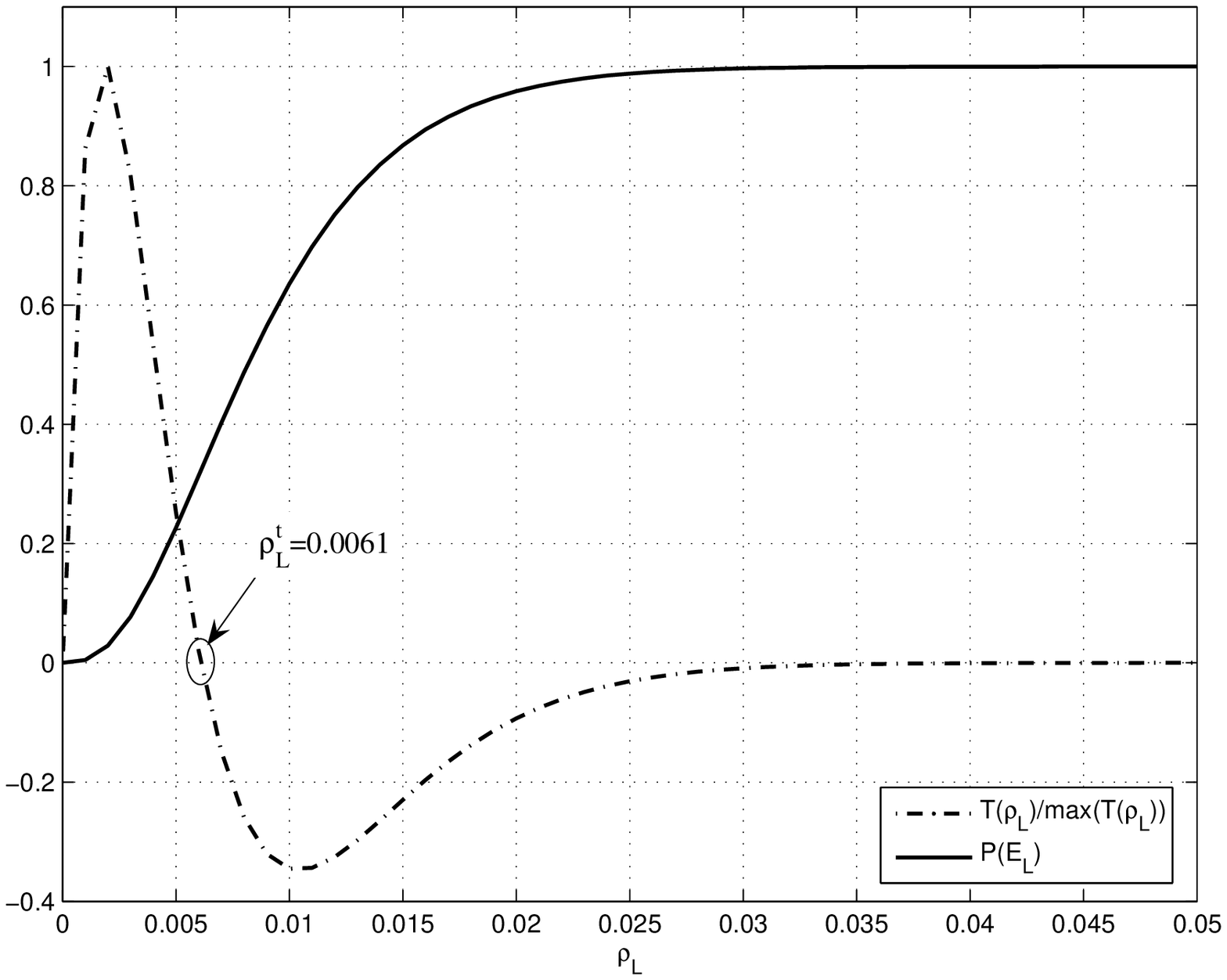}{Behavior of the localization
probability $P(E_L)$ as a function of the L-node density $\rho_L$.
Other transmission parameters are $\beta_{th}=40$~dB,
$\sigma_{s}=4$~dB, $n_p=2$, $\rho_{NL}=0.1$ NL-nodes$/m^2$ and
$R=100$m.}{Esempio_thr_ro_L_pel}
Returning to our analysis where we assume the knowledge of the
radio coverage area of a given NL-node, a common characteristic of
many problems tackled using the probabilistic method is the
existence of transition thresholds where the event of interest
exhibits a large variation. Indeed, it is known that
every monotone graph property in randomly generated graphs has a
sharp transition threshold \cite{Friedgut,GoelSanatan}. Such
thresholds are established in the asymptotic case, i.e., in the
limit when the number of nodes in the random graph tends to
infinity. Thresholds are very useful in practice for topology
control of the network \cite{SantiTesto}.

In what follows, we will first derive transition thresholds for
the localization problem in finite regimes, i.e., when the numbers
of both L and NL-nodes are finite within a bounded domain $S$ as
defined in the previous sections. In the second part, we will
investigate the localization problem in the limiting cases of
dense networks. Notice that our results hold even in the random
geometric model by setting $\sigma_s=0$.
\subsection{Thresholds for Single Node Localization Probability, Finite Case}
Since the localization probability $P(E_L)$ in
(\ref{loc_probability_R2_2}) is a monotonically increasing
function of its arguments embraced within $\lambda_{NL}$, the
transition thresholds observable in the finite regime (especially
for large values of $\rho_L$) can be obtained by taking the second
partial derivative of $P\left(E_L\right)$ in
(\ref{loc_probability_R2_2}) with respect to the parameters of
interest, such as $\rho_L$ and $d_{max}$, and setting the result
to zero.

Let S be the usual bounded circular domain of radius $R$ in
$\Re^2$. Let us analyze the thresholds of $P\left(E_L\right)$ with
respect to $\rho_L$. Let $\gamma_1=\pi d_{max}^2
e^{\frac{\eta^2}{\alpha^2}}$. After some algebra, the first
partial derivative with respect to $\rho_L$ can be expressed as
\begin{equation}\label{Loc_prob_thresh_1_pel}
\frac{\partial
}{\partial\rho_L}P\left(E_L\right)=e^{-\gamma_1\rho_L}\left[\frac{1}{2}\gamma_1^3\rho_L^2\right]
\end{equation}
Given that $\rho_L>0$, (\ref{Loc_prob_thresh_1_pel}) is always
greater than zero, showing a strictly increasing behavior of
$P\left(E_L\right)$ with respect to $\rho_L$.

The second partial derivative $T(\rho_L)=\frac{\partial^2
}{\partial\rho_L^2}P\left(E_L\right)$ of $P\left(E_L\right)$ with
respect to $\rho_L$ is:
\begin{equation}\label{Loc_prob_thresh_1_2_pel}
%
T(\rho_L)=e^{-\gamma_1\rho_L}\gamma_1^3\rho_L\left[1-\frac{\gamma_1}{2}\rho_L\right]
%
\end{equation}
The values of the threshold $\rho^t_L$ are the solutions of the
equation $T(\rho_L)=0$, that is,
\figura{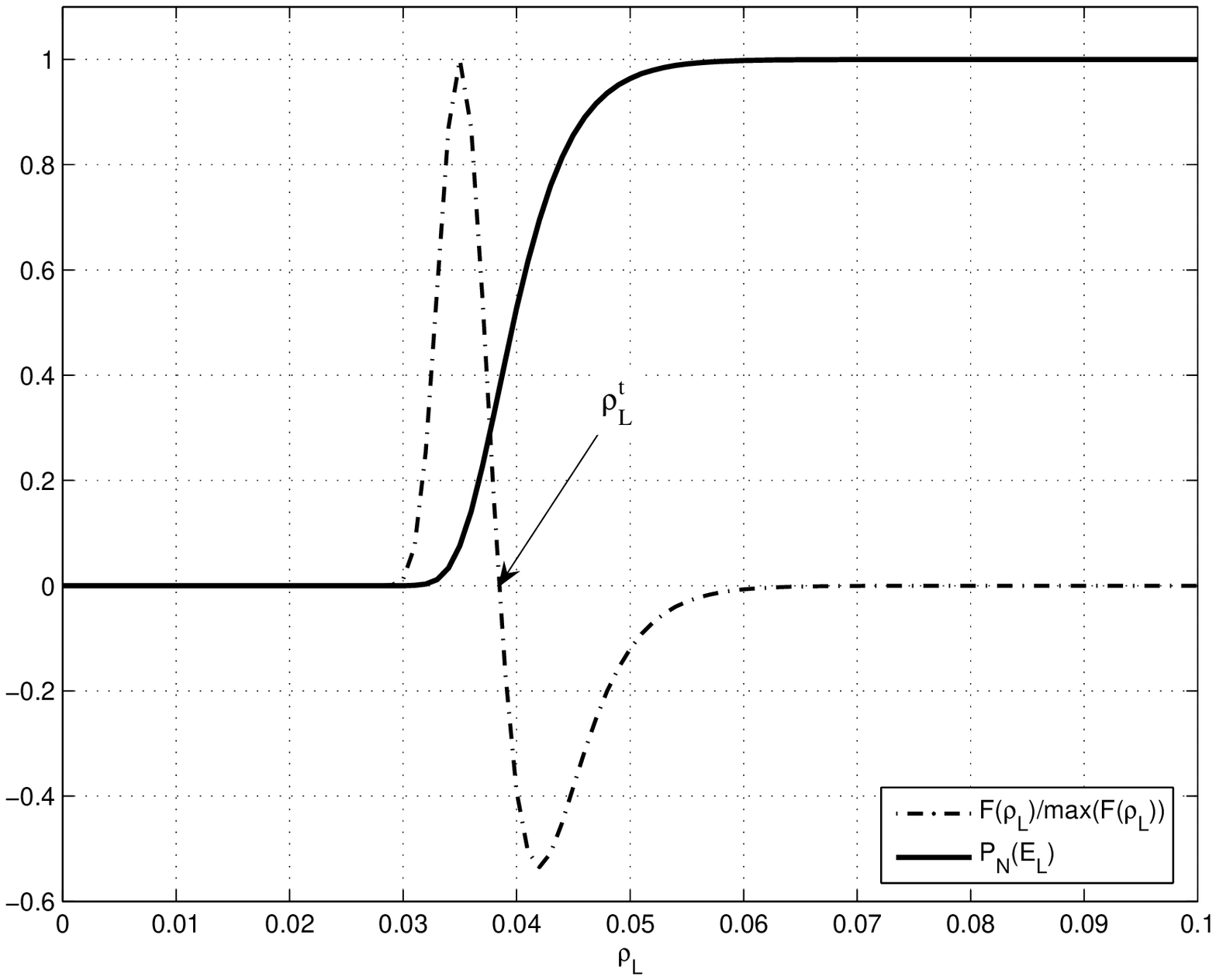}{Behavior of the localization
probability $P_N(E_L)$ as a function of the L-node density
$\rho_L$. Other transmission parameters are $\beta_{th}=40$~dB,
$\sigma_{s}=4$~dB, $n_p=2$, $\rho_{NL}=0.1$ NL-nodes$/m^2$ and
$R=100$m.}{Esempio_thr_ro_L}
\begin{equation}\label{Thresh_Loc_prob_thresh_1_2_pel}
1-\frac{\gamma_1}{2}\rho_L=0\Rightarrow\rho^t_L=\frac{2}{\pi
d^2_{max}}e^{-\frac{\eta^2}{\alpha^2}}
\end{equation}
Fig.~\ref{Esempio_thr_ro_L_pel}, shows the behavior of the
localization probability $P(E_L)$ as a function of $\rho_L$ for
the transmission setup noted in the figure caption. Moreover, in
the figure we report the behavior of the second derivative
$T(\rho_L)$ (normalized with respect to its maximum for depicting
both curve on the same ordinate range) along with the threshold
$\rho^t_L$ obtained by (\ref{Thresh_Loc_prob_thresh_1_2_pel}) with
the setup noted above.

Let us analyze the thresholds of $P\left(E_L\right)$ with respect
to $d_{max}$, and for ease of notation, set $d_{max}=d_m$ and
$\gamma_2=\rho_L\pi e^{\frac{\eta^2}{\alpha^2}}$. Following the same
reasoning as applied for $\rho^t_L$, after some algebra, one easily
obtains the threshold for the localization probability with
respect to the node transmission range $d_m$:
\begin{equation}\label{Thresh_Loc_prob_thresh_dmax_1_2_pel}
d_{m}^t=\sqrt{\frac{2}{\pi \rho_L}}e^{-\frac{\eta^2}{2\alpha^2}}
\end{equation}
\subsection{Thresholds for the Localization Probability of the Whole Network of NL-nodes, Finite Case}
Owing to the fact that $P_N(E_L)\le P(E_L)$
for a given transmission scenario, thresholds for the probability
$P_N(E_L)$ are expected to be higher than the ones obtained for
$P(E_L)$.

Let us start our analysis by deriving the thresholds of
$P_N\left(E_L\right)$ in (\ref{loc_probability_allnetw0}) with
respect to $\rho_L$. Let $\gamma_1=\pi d_{max}^2
e^{\frac{\eta^2}{\alpha^2}}$. After some algebra, the second
partial derivative $F(\rho_L)=\frac{\partial^2
}{\partial\rho_L^2}P_N\left(E_L\right)$ of $P_N\left(E_L\right)$
with respect to $\rho_L$ is:
\begin{equation}\label{Loc_prob_thresh_1}
\begin{array}{ll}
F(\rho_L)=\frac{1}{2}\gamma_1^2N_{NL}\left[1-e^{-\gamma_1\rho_L}\left(1+\gamma_1\rho_L+\frac{1}{2}\gamma_1^2\rho_L^2\right)\right]^{N_{NL}-1}&\\
\cdot\left[e^{-\gamma_1\rho_L}(2\rho_L-\gamma_1\rho_L^2)+\frac{1}{2}\frac{\gamma_1^3\rho_L^4e^{-2\gamma_1\rho_L}\left(N_{NL}-1\right)}{1-e^{-\gamma_1\rho_L}\left(1+\gamma_1\rho_L+\frac{1}{2}\gamma_1^2\rho_L^2\right)}\right]&
\end{array}
\end{equation}
The values of the threshold $\rho^t_L$ are the solutions of the
equation $F(\rho_L)=0$. Noting that
$$e^{+\gamma_1\rho_L}>\left(1+\gamma_1\rho_L+\frac{1}{2}\gamma_1^2\rho_L^2\right)$$
with $\rho_L>0$ and $\gamma_1>0$, the only solutions are
the roots of the non-linear equation:
\figura{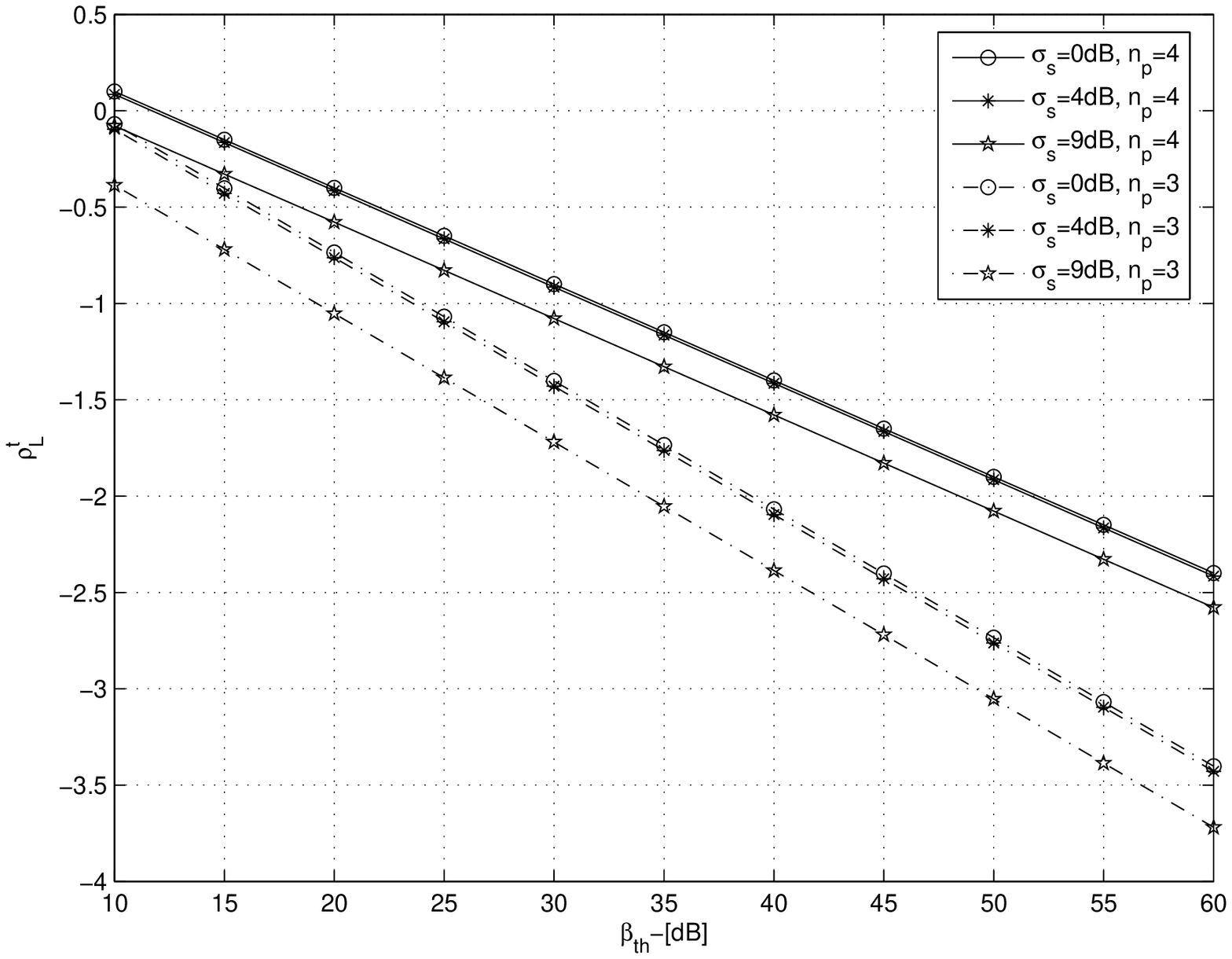}{Finite case thresholds
$\log_{10}(\rho^t_L)$ for $P_N(E_L)$ as a function of $\beta_{th}$
for a variety of parameters noted in the legend. Other
transmission parameters common to all plots are $\rho_{NL}=0.1$
NL-nodes$/m^2$ and $R=100$m.}{finite_thr_ro_L}
\begin{equation}\label{thresholds_PNE_L_rhoL}
2-\gamma_1\rho_L+\frac{1}{2}\frac{\gamma_1^3\rho_L^3e^{-\gamma_1\rho_L}\left(N_{NL}-1\right)}{1-e^{-\gamma_1\rho_L}\left(1+\gamma_1\rho_L+\frac{1}{2}\gamma_1^2\rho_L^2\right)}=0
\end{equation}
As a reference example, consider the transmission scenario
investigated in the previous section, and summarized in the
caption of Fig.~\ref{Esempio_thr_ro_L} which shows the behavior of
the localization probability $P_N(E_L)$ as a function of $\rho_L$.
Also shown is the behavior of the second derivative $F(\rho_L)$
(normalized with respect to its maximum for depicting both curve
on the same ordinate range). Note that the threshold for
$P_N(E_L)$ is about one order of magnitude greater than the
threshold $\rho^t_L$, noted
in~(\ref{Thresh_Loc_prob_thresh_1_2_pel}), relative to $P(E_L)$.

The behavior of the thresholds (obtained as the solutions
of~(\ref{thresholds_PNE_L_rhoL})) as a function of the parameter
$\beta_{th}$ for various values of the path-loss exponent $n_p$
and $\sigma_s$ is depicted in Fig.~\ref{finite_thr_ro_L}. From
this figure, we observe the decreasing behavior of
$\rho^t_L$ for increasing values of $\beta_{th}$, i.e. for
increasing values of the maximum transmission range $d_{max}$
noted in (\ref{equat_d_max}).

Let us analyze the thresholds of $P_N\left(E_L\right)$ with
respect to $d_{max}$, and for ease of notation, set $d_{max}=d_m$.
Let $\gamma_2=\rho_L\pi e^{\frac{\eta^2}{\alpha^2}}$. After some
algebra, the second partial derivative $F(\rho_L)=\frac{\partial^2
}{\partial d_{m}^2}P_N\left(E_L\right)$ of $P_N\left(E_L\right)$
with respect to $d_{m}$ is:
\figura{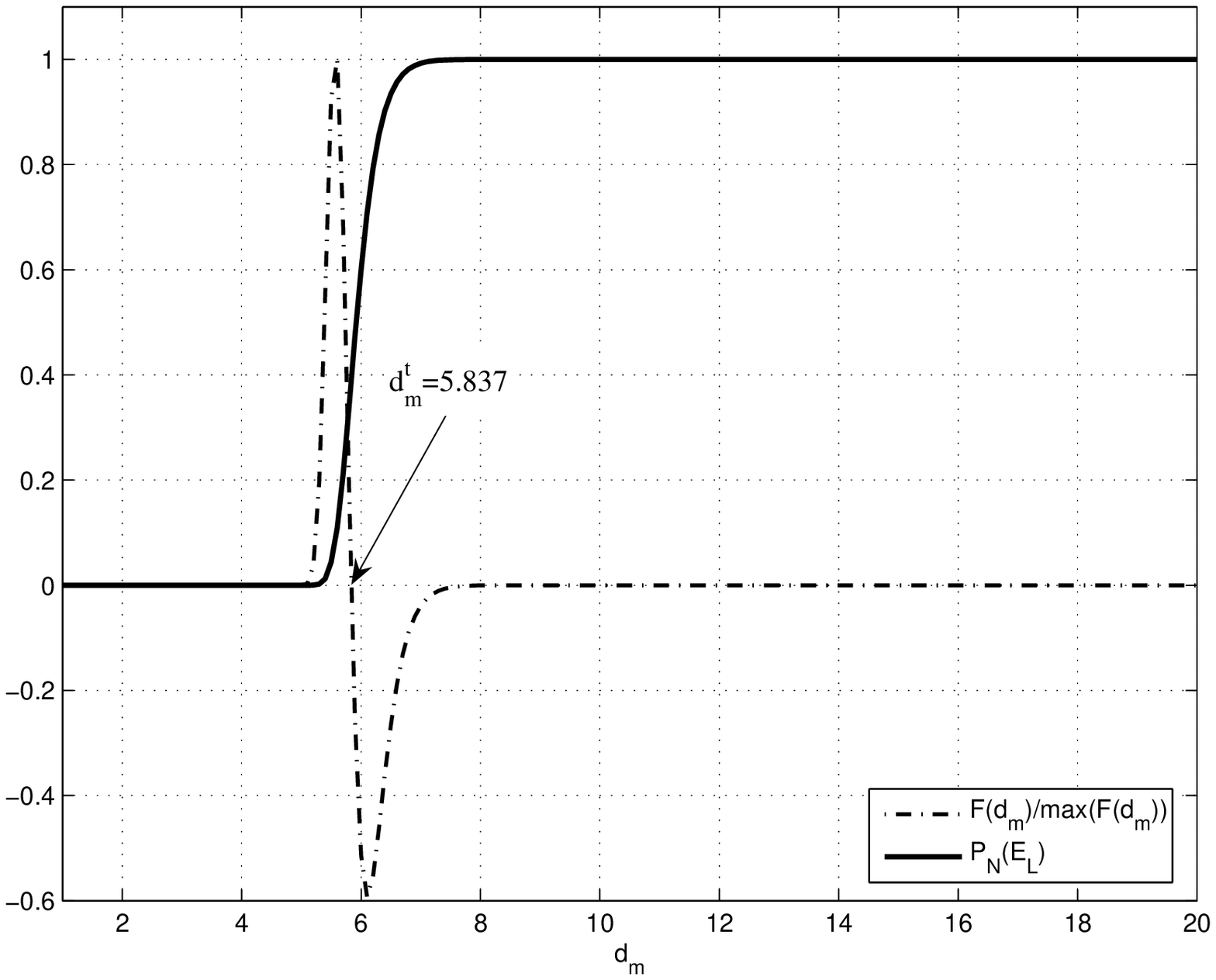}{Behavior of the localization
probability $P_N(E_L)$ as a function of the maximum transmission
range $d_{max}$. Other transmission parameters are
$\beta_{th}=40$~dB, $\sigma_{s}=4$~dB, $n_p=2$,
$\rho_{L}=\rho_{NL}=0.1$ NL-nodes$/m^2$ and
$R=100$m.}{Esempio_thr_d_max}
\begin{equation}\label{Loc_prob_thresh_2}
\begin{array}{ll}
F(d_{m})=\left[1-e^{-\gamma_2d_{m}^2}\left(1+\gamma_2d_{m}^2+\frac{1}{2}\gamma_2^2d_{m}^4\right)\right]^{N_{NL}-1}\cdot &\\
\gamma_2^3N_{NL}d_{m}^4e^{-\gamma_2d_{m}^2}\cdot\left[5-2\gamma_2d_{m}^2+\frac{\gamma_2^3d_{m}^6e^{-\gamma_2d_{m}^2}\left(N_{NL}-1\right)}{1-e^{-\gamma_2d_{m}^2}\left(1+\gamma_2d_{m}^2+\frac{1}{2}\gamma_2^2d_{m}^4\right)}\right]&
\end{array}
\end{equation}
The values of the threshold $d_m^t$ are the solutions of the
equation $F(d_{m})=0$. Upon noting that
$$e^{+\gamma_2d_{m}^2}>\left(1+\gamma_2d_{m}^2+\frac{1}{2}\gamma_2^2d_{m}^4\right),\forall d_m>0,\gamma_2>0$$
the only solutions are the roots of the non-linear
equation:
\begin{equation}\label{Thr_finite_PNEL_d_m}
5-2\gamma_2d_{m}^2+\frac{\gamma_2^3d_{m}^6e^{-\gamma_2d_{m}^2}\left(N_{NL}-1\right)}{1-e^{-\gamma_2d_{m}^2}\left(1+\gamma_2d_{m}^2+\frac{1}{2}\gamma_2^2d_{m}^4\right)}=0
\end{equation}
Fig.~\ref{Esempio_thr_d_max} shows the behavior of the network
localization probability $P_N(E_L)$ as a function of $d_{m}$ for
the transmission setup noted in the figure caption. The figure
also shows the behavior of the second derivative $F(d_m)$
(normalized with respect to its maximum for depicting both curve
on the same ordinate range) along with the threshold $d^t_m$
obtained by solving the non-linear equation
(\ref{Thr_finite_PNEL_d_m}) with the setup noted in the caption of
Fig.~\ref{Esempio_thr_d_max}.

The behavior of the thresholds (obtained as the solutions
of~(\ref{Thr_finite_PNEL_d_m})) as a function of the L-node
density $\rho_{L}$ for $n_p=4$ and various values of $\sigma_s$ is
depicted in Fig.~\ref{finite_thr_d_max}. From this figure, we observe
the decreasing behavior of $d^t_m$ for increasing
values of $\rho_{L}$.
\figura{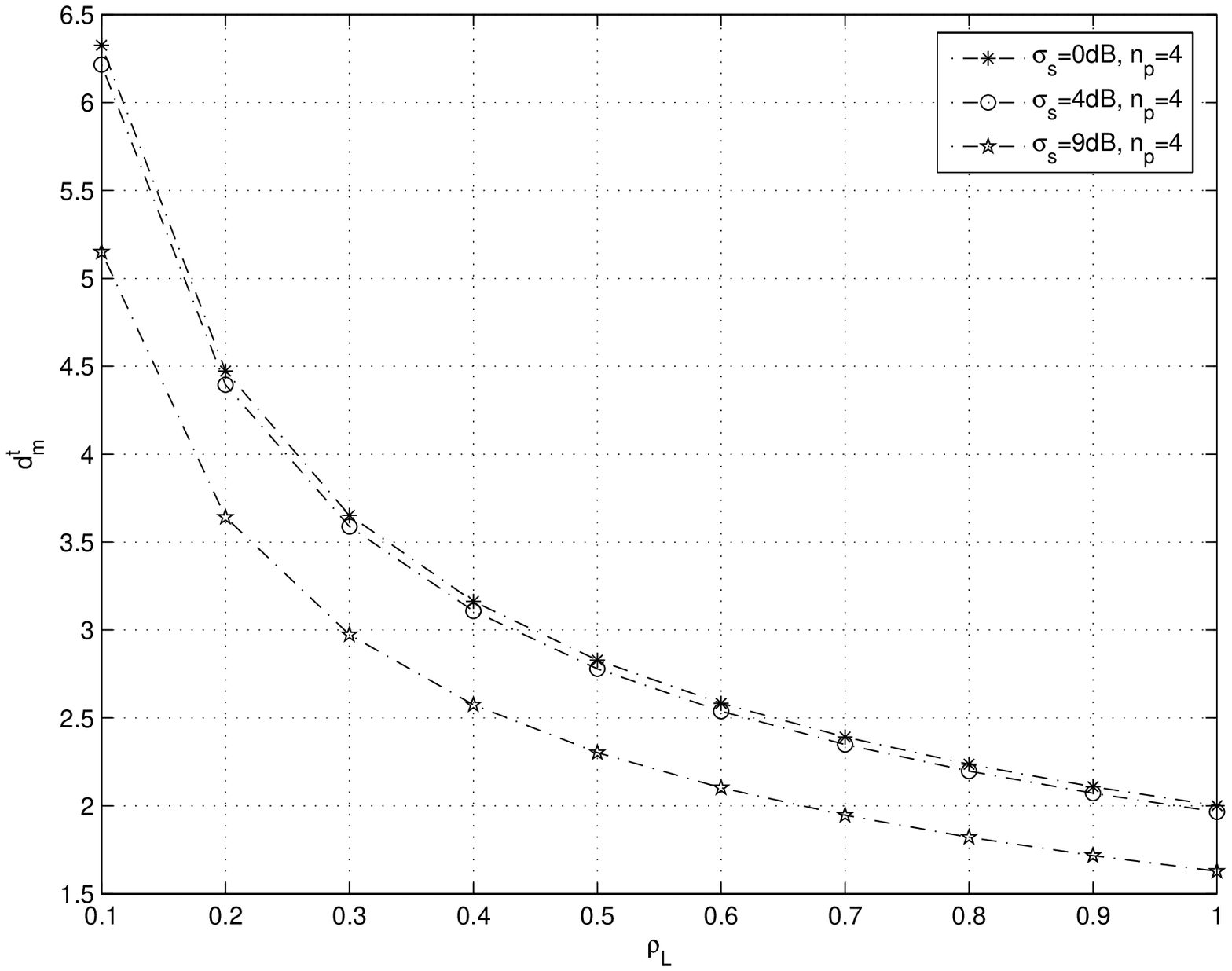}{Finite case thresholds $d^t_m$ of the
localization probability $P_N(E_L)$ as a function of the L-node
density $\rho_L$ for a variety of parameters noted in the legend.
Other transmission parameters are $\rho_{NL}=0.1$ NL-nodes$/m^2$
and $R=100$m.}{finite_thr_d_max}
\section{Asymptotic Behavior of the Localization Probability and Thresholds}
\label{Section_Asymptotic_Behaviour}
In this section, we present results on the behavior of the
localization probabilities of both single NL-node and the overall
network of NL-nodes deployed over both bounded and unbounded
domains in a transmitting scenario affected by shadow fading.

The first result concerns dense networks, i.e., network of nodes
whereby the node densities of both point processes deployed over a disk
$S\subset\Re^2$ with radius $R\gg d_{max}$,
are allowed to grow unboundedly as a function of the number of
nodes over $S$. As above, edge effects are neglected, and the
hypothesis $R\gg d_{max}$ allows us to employ the relation
$\lambda_{NL,R}\approx\lambda_{NL}$. Moreover, assume that the
transmission range is homogeneous and equal to $d_{max}$ for both
kinds of nodes. The next theorem investigates the behavior of the
localization probability $P_N(E_L)$ of the network over $S$ in
terms of the orders of growth
of the number of L and NL-nodes over $S$.\\

\noindent\textbf{Theorem 1 (dense networks).} Let S be a bounded
disk of radius $R$ belonging to $\Re^2$. Assume that two sets of
nodes with statistically independent Poisson point processes with
densities $\rho_L$ and $\rho_{NL}$ are deployed over $S\subseteq
\Re^2$. Let $N_L$ and $N_{NL}$ be the number of L-nodes and
NL-nodes, respectively, falling in $S$, and assume that $N_L$ and
$N_{NL}$ asymptotically grow as the functions $f_L(n)$ and
$f_{NL}(n)$, where $n$ is an asymptotic growth parameter.

The network of NL-nodes gets a.a.s. localized, i.e.,
\[
\lim_{n\rightarrow\infty}P_N(E_L)=1
\]
for any $f_L(n)$ and $f_{NL}(n)$ such that
\[
\lim_{n\rightarrow\infty} f_{NL}(n)f^2_L(n)e^{-\gamma f_L(n)}=0
\]
whereby $\gamma$ is an appropriate real constant greater than zero.\\

\noindent\textbf{Proof.}
Consider $P_N\left(E_L\right)$ in~(\ref{loc_probability_allnetw0})
along with the relation~(\ref{temp_prob0}), and the following
inequalities~\cite{Widder}:
\begin{eqnarray}\label{approximations}
(1+x)^n &<& e^{nx},~\forall x\in \Re,~x\ne 0\\
1-xy &\le& (1-x)^y,~0<x\le 1\le y
\end{eqnarray}
Based on the previous two relations, $P_N\left(E_L\right)$ in
(\ref{loc_probability_allnetw0}) can be bounded as follows:
\begin{equation}\label{loc_probability_allnetw_bound}
1-X(\lambda_{NL})\cdot N_{NL}\le P_N\left(E_L\right)<
e^{-X(\lambda_{NL})\cdot N_{NL}}
\end{equation}
where, $N_{NL}\ge 1$ and $X(\lambda_{NL})\le 1$ by definition.
Equ. (\ref{loc_probability_allnetw_bound}) will be used for
demonstrating the three claims of the theorem.

It suffices to demonstrate that as $n\rightarrow\infty$,
$X(\lambda_{NL})\cdot N_{NL}\rightarrow 0$ so that
$P_N\left(E_L\right)\rightarrow 1$, i.e., the network of NL-nodes
over $S$ gets localized w.h.p.

Let us rewrite $X(\lambda_{NL})$ in an appropriate form for
successive developments. Upon setting
\begin{equation}\label{def_gamma}
\begin{array}{lll}
\gamma&=&\left(\frac{d_{max}}{R}\right)^2
e^{\frac{\eta^2}{\alpha^2}}\\
N_L &=& \rho_L \pi R^2
\end{array}
\end{equation}
$X(\lambda_{NL})\cdot N_{NL}$ can be rewritten as follows:
\begin{equation}\label{X_temp}
\begin{array}{lll}
X(\lambda_{NL})\cdot N_{NL}&=&N_{NL}e^{-\gamma N_L}\left[1+\gamma
N_L+\frac{1}{2}\gamma^2 N_L^2\right]\\
&=&c\cdot N_{NL}N_L^2e^{-\gamma N_L} 
\end{array}
\end{equation}
whereby
$c=\left[\frac{1}{N_L^2}+\frac{\gamma}{N_L}+\frac{\gamma^2}{2}\right]$.

From (\ref{X_temp}), it is straightforward to demonstrate that for
any $f_L(n)$ and $f_{NL}(n)$ such that
\[
\lim_{n\rightarrow\infty} N_{NL}N_L^2e^{-\gamma
N_L}=\lim_{n\rightarrow\infty} f_{NL}(n)f^2_L(n)e^{-\gamma
f_L(n)}=0
\]
the network of NL-nodes over $S$ gets localized a.a.s.
\begin{flushright}$\Box$\end{flushright}
The previous theorem is the starting point for identifying
appropriate orders of growth of both L and NL-nodes guaranteeing
asymptotically almost sure localization. In this respect, we note
the following corollary.\\

\noindent\textbf{Corollary (dense networks).} Under the scenario
described in Theorem 1, as $n\rightarrow\infty$ the following
holds:
\begin{enumerate}
    \item Suppose $N_{NL}\sim f_{NL}(n)\sim q\cdot n^{1-\xi}$ with $\xi\in [0,1)$ and
    $N_L\sim f_L(n)\sim p\cdot\ln (n)$, with $p$ and $q$ two suitable constants strictly greater
than zero.

Then, the network of NL-nodes over $S$ gets localized w.h.p. as
$n\rightarrow\infty$ provided that
$$p>p_0=\left(\frac{R}{d_{max}}\right)^2 \left(1-\xi\right)
e^{-\frac{\eta^2}{\alpha^2}}$$
    \item Suppose $N_{L}\sim f_{L}(n)\sim \ln \left(f_{NL}(n)\right)$.

Then, the network of NL-nodes over $S$ gets localized w.h.p. as
$n\rightarrow\infty$ provided that
$$\left(\frac{d_{max}}{R}\right)^2 e^{\frac{\eta^2}{\alpha^2}}>1$$
\item Suppose $N_L\sim f_{L}(n)\sim n$ and $N_{NL}\sim
f_{NL}(n)\sim n^t$ with $t > 0$ as $n\rightarrow\infty$. Then, the
network of NL-nodes over $S$ gets localized w.h.p. as
$n\rightarrow\infty$.

    \item As a consequence of the previous point, suppose $N_{NL}= f_{NL}(n)\sim O(1)$, that is, $N_{NL}$ is a bounded sequence.
    Then, the network of NL-nodes over $S$ gets localized w.h.p. as
$n\rightarrow\infty$ provided that $N_L\sim f_L(n)\sim \omega(n)$
with $\omega(n)\rightarrow\infty$ no matter how slowly $\omega(n)$
grows.

\end{enumerate}

\noindent\textbf{Proof.}
As far as claim 1) of the corollary is concerned, it suffices to
demonstrate that as $n\rightarrow\infty$, $X(\lambda_{NL})\cdot
N_{NL}\rightarrow 0$ for $N_{NL}\sim q n^{1-\xi}$ with $\xi\in
[0,1)$ and $N_L\sim p\ln (n)$ with $p$ and $q$ two suitable
constants strictly greater than zero.

If $N_L\sim p\ln (n)+o(\ln (n))$ with $p$ a suitable constant
$p>0$, it follows that,
\begin{equation}\label{X_temp_asymp_1}
\begin{array}{lll}
X(\lambda_{NL})\cdot N_{NL}&\sim& c\cdot p^2 \ln^2(n)N_{NL}\cdot
e^{-\gamma p
\ln(n)}\\
&=&c\cdot p^2 \ln^2(n)N_{NL}\cdot n^{-\gamma p}
\end{array}
\end{equation}
In the case $N_{NL}\sim q n^{1-\xi}$ with $\xi\in [0,1)$, for
$n\rightarrow\infty$ we have:
\begin{equation}\label{X_temp_asymp_2}
\begin{array}{lll}
X(\lambda_{NL})\cdot N_{NL}&\sim& c\cdot q\cdot p^2 \ln^2(n)\cdot
n^{1-\xi-\gamma p}
\end{array}
\end{equation}
When $n\rightarrow\infty$, $X(\lambda_{NL})\cdot N_{NL}\rightarrow
0$ if the following relation holds:
\[
1-\xi-\gamma p<0
\]
since we have \cite{Widder},
\[
\lim_{x\rightarrow\infty}\frac{\left[\ln(x)\right]^\alpha}{x^\beta}=0,~\forall~
\alpha,\beta>0
\]
By substituting the definition of $\gamma$ in the previous
relation, after some algebra the following threshold follows:
\begin{equation}\label{threshold_1}
p>p_0=\left(\frac{R}{d_{max}}\right)^2 \left(1-\xi\right)
e^{-\frac{\eta^2}{\alpha^2}}
\end{equation}
Claim 2) follows from observing that for $N_{L}\sim f_{L}(n)\sim
\ln \left(f_{NL}(n)\right)$, (\ref{X_temp}) can be rewritten as
\begin{equation}\label{X_temp_asymp_3claim2}
\begin{array}{lll}
X(\lambda_{NL})\cdot N_{NL}&\sim&
\left(f_{NL}(n)\right)^{1-\gamma}\ln^2\left(f_{NL}(n)\right)
\end{array}
\end{equation}
As $n\rightarrow\infty$, it is
\[
\lim_{n\rightarrow\infty}\left(f_{NL}(n)\right)^{1-\gamma}\ln^2\left(f_{NL}(n)\right)=0
\]
provided that $1-\gamma<0$, from which
\[
\gamma=\left(\frac{d_{max}}{R}\right)^2
e^{\frac{\eta^2}{\alpha^2}}>1
\]

Claim 3) follows from observing that for $N_L\sim n$ and
$N_{NL}\sim n^t$, the following holds;
\begin{equation}\label{X_temp_asymp_3}
\begin{array}{lll}
X(\lambda_{NL})\cdot N_{NL}&\sim& c \cdot n^{2+ t}e^{-\gamma
n}\rightarrow 0,~ n\rightarrow\infty
\end{array}
\end{equation}
no matter what the order $t$ of growth of the number of NL-nodes.
So, asymptotically, the network of NL-nodes gets always localized
w.h.p. under these conditions.

Finally, claim 4) follows from the proof of claim 1) upon
considering $\xi=1$ in (\ref{X_temp}). Note that based on the
proof of claim 1), $\xi=1$ signifies the fact that $N_{NL}=O(1)$,
i.e., $N_{NL}$ is a bounded sequence, and that
$X(\lambda_{NL})N_{NL}\sim N_L^2e^{-\gamma N_L}\rightarrow 0$ for
any $N_L\sim \omega(n)\rightarrow\infty$ as $n\rightarrow\infty$.
\begin{flushright}$\Box$\end{flushright}
Since inequality (\ref{threshold_1}) in Claim 1) is the most
important result of this corollary, some considerations are in
order. The basic meaning of this result is as follows; in a
bounded circular region $S\subset \Re^2$ with area $\pi R^2$ with
$R\gg d_{max}$, the network of randomly deployed NL-nodes gets
asymptotically localized even though the number of L-nodes grows
only logarithmically (i.e., with an order of growth smaller than
that of the NL-nodes) provided that the constant $p$ is above the
threshold $p_0$.
%
%
This result is fundamental from a point of view of network
topology, since it assures us that a number of L-nodes which grows
only logarithmically suffice for assuring network localization,
provided that $p>p_0$, even though the number of NL-nodes grows
faster than logarithmically. It is worth noting that these results
also hold for random geometric graphs (RGG); in a transmission
scenario typical of RGGs, whereby any NL-node can communicate with
any other L-node within the distance $r=d_{max}$, we have
$\sigma_s=0$ ($\Rightarrow\eta=0$), and the threshold becomes:
\figura{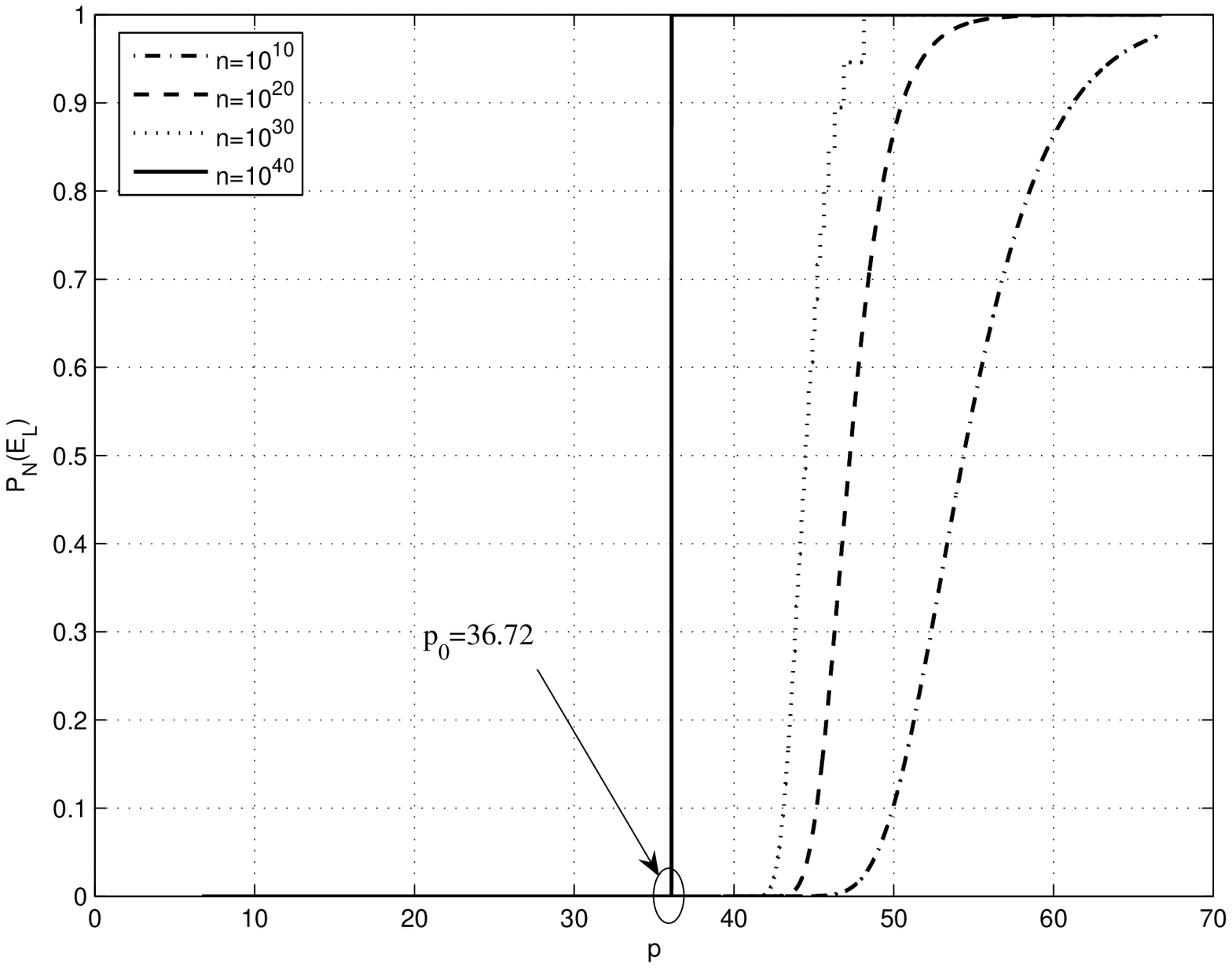}{Behavior of the localization
probability $P_N(E_L)$ as a function of the constant $p$ in
$N_L\sim p\cdot \log (n)$ for unboundedly increasing values of
$n$. Transmission scenario is compliant with the following
parameters; $\xi=0.51$ ($N_{NL}\sim n^{1-\xi}$), $\sigma_s=9$ dB,
$n_p=4$, $\beta_{th}=30$ dB, $R=60$ m, and $d_{max}\approx 5.62$
m. With this setup, the threshold $p_0=36.72$.}{asympt_thr_p_0}
\[
p_{0,RGG}=\left(\frac{R}{d_{max}}\right)^2 \left(1-\xi\right)
\]
Borrowing the terminology used in the context of random graph
theory \cite{palmer}, claim 1) of the previous corollary states
that the function $N_L\sim p_0\ln (n)+o(\ln(n))$ is a threshold
for the localization problem at hand. Any function $N_L\sim
o(p_0\ln (n))$ allows network localization asymptotically w.h.p.

As a reference example, Fig. \ref{asympt_thr_p_0} shows the
behavior of the localization probability $P_N(E_L)$ as a function
of $p$ for unboundedly increasing values of $n$ in the
transmitting scenario summarized in the figure caption. Note that,
for $p<p_0=36.72$, $P_N(E_L)$ is always zero, while $P_N(E_L)$
becomes instantaneously unitary so long as $p=p_0$ while
$n\rightarrow\infty$.

Finally, notice that such a threshold does not hold for single
NL-node localization probability. In other words, upon considering
the probability $P(E_L)$ in (\ref{loc_probability_R2_2}) for
single NL-node probability, it is simple to observe that any
randomly chosen NL-node over a bounded domain $S$ gets localized
w.h.p. for $N_L\sim \omega (n)$, whatever the behavior of the
function $\omega (n)$, provided that $\omega (n)\rightarrow\infty$
as $n\rightarrow\infty$.

The results obtained for dense networks state conditions for
a.a.s. localization of a network of NL-nodes over a bounded
circular domain for a variety of orders of growth of the number of
NL-nodes deployed.

Let us now look at the problem from a different perspective. In
other words, we look at the problem by considering constant L-node
density while we let the size of the domain $S$ to grow in such a
way that $\rho_{L}=\frac{N_{L}}{\pi R^2}=O(1)$. Such a result is
typical of non-dense networks. In this respect, it is useful to
evaluate the
minimum $d_{max}$ above which the network of NL-nodes gets localized a.a.s.\\

\noindent\textbf{Theorem 2 (unbounded domains, constant
densities).}
Let S be a disk of radius $R$ belonging to $\Re^2$. Assume that
two sets of nodes with statistically independent Poisson point
processes with densities $\rho_L$ and $\rho_{NL}$ are deployed
over $S\subseteq \Re^2$. Let $N_L$ and $N_{NL}$ be, respectively,
the number of L-nodes and NL-nodes falling in $S$, and consider
any asymptotically increasing function $\omega(n)$, such that
$\omega(n)\rightarrow\infty$ as $n\rightarrow\infty$, and assume
that $N_{NL}\sim o(\omega^{-2}(n)e^{+\omega (n)})$.
%

Moreover, assume that, as $R\rightarrow\infty$, the L-node density
satisfies the following relation:
\begin{equation}\label{condizioni_su_ro}
\begin{array}{ll}
\rho_{L}=\frac{N_{L}}{\pi R^2}=O(1)&
\end{array}
\end{equation}
Then, as $n\rightarrow\infty$ in such a way that
(\ref{condizioni_su_ro}) holds, the network of NL-nodes gets
a.a.s. localized if,
\begin{equation}\label{equazionedi_dmax}
d_{max}=\sqrt{\frac{e^{-\frac{\eta^2}{\alpha^2}}}{\pi\rho_L}\omega(n)}
\end{equation}
\noindent\textbf{Proof.} The proof follows an outline similar to
the one of the previous theorem. Consider $P_N(E_L)$
in~(\ref{loc_probability_allnetw0}) along with its bound in
(\ref{loc_probability_allnetw_bound}). As before, the objective is
to show that asymptotically, the transmission range $d_{max}$
between each pair of L-NL-nodes should grow at least as specified
in (\ref{equazionedi_dmax}) in order for $P_N(E_L)\rightarrow 1$
as $n\rightarrow\infty$.

Given $ N_{NL}$, $X(\lambda_{NL})N_{NL}$ can be rewritten as
follows:
\begin{equation}\label{temp_prob_dmax}
\begin{array}{lll}
X(\lambda_{NL})N_{NL}&=&N_{NL} e^{-\rho_L\pi
d^2_{max}e^{\frac{\eta^2}{\alpha^2}}}\left[1+\right.\\
&&+\rho_L\pi
d^2_{max}e^{\frac{\eta^2}{\alpha^2}}+\left.\frac{\rho_L^2}{2}\pi^2
d^4_{max}e^{2\frac{\eta^2}{\alpha^2}}\right]
\end{array}
\end{equation}
With this setup and given (\ref{loc_probability_allnetw_bound}), it
suffices to show that $X(\lambda_{NL})N_{NL}\rightarrow 0$ when
$d_{max}$ grows as stated in (\ref{equazionedi_dmax}).

Upon substituting $d_{max}$ given in (\ref{equazionedi_dmax}) in
(\ref{temp_prob_dmax}), the following relation follows:
\[
\begin{array}{lll}
X(\lambda_{NL})N_{NL}&=&N_{NL}\cdot
 e^{-\omega(n)}\left[1+\omega(n)+\frac{1}{2}\omega^2(n)\right]\\
 &\sim&\frac{1}{2} N_{NL}\cdot
 \omega^2(n) e^{-\omega(n)}
\end{array}
\]
which goes to zero so long as $\omega(n)\rightarrow\infty$ as
$n\rightarrow\infty$ for any
$N_{NL}=o(\omega^{-2}(n)e^{\omega(n)})$, guaranteeing that the
network of NL-nodes gets localized w.h.p.
\begin{flushright}$\Box$\end{flushright}
The result stated in this theorem is reminiscent of percolation
theory. In other words, when the deployment region $S$ tends to
become the entire plane $\Re^2$ (i.e., $R\rightarrow\infty$)
in such a way that $\rho_L$ is a finite and
constant value, the entire network of NL-nodes becomes a giant
localized component so long as the transmission range $d_{max}$
takes on the values expressed by (\ref{equazionedi_dmax}) provided
that $N_{NL}=o(\omega^{-2}(n)e^{\omega(n)})$.

As an example, if $\omega (n)\sim \ln (n)$, and
$$N_{NL}\sim o\left(\frac{n}{\ln^2n}\right)$$
the network with an ever-increasing size gets asymptotically
localized so far as $d^2_{max}$ grows at least as $d^2_{max}\sim
\ln n$.

Notice that, since in practice no real device can support an
ever-increasing communication range $d_{max}$, as the network
domain increases in size, in the limit there is always a non-zero
probability that some node cannot get localized.
\section{Conclusions}
\label{Section_Conclusions}
The aim of this paper has been manyfold. Considering a two
dimensional domain $S\subseteq\Re^2$ over which two sets of nodes
following statistically independent uniform Poisson point
processes with constant densities $\rho_L$ and $\rho_{NL}$ are
deployed, we first derived the probability that a randomly chosen
NL-node over $S$ gets localized as a function of a variety of
system level parameters. Then, we investigated the probability
that the whole network of NL-nodes over $S$ gets localized. The
transmission scenario assumed is that of shadow fading.

Furthermore, we presented a theoretical framework for
deriving both finite case and asymptotic thresholds for the probability
of localization in connection with both a single non-localized node
randomly chosen over the investigated domain, and the whole
network of non-localized nodes. Finally, we investigated the
presence of thresholds on the problem at hand for unboundedly
increasing values of the number of deployed nodes over the domain
$S$.
\section*{Appendix I}
Upon substituting (\ref{P_ji_definitiva}) in
(\ref{num_neighbors}), and considering $r=\overline{d}_{j,i}$:
%
%
\begin{eqnarray}\label{grado_medio_dv}
E\{d^{NL}_v|R\}&=&2\pi
\rho_L\int_{0}^{R}\frac{1}{2}\left[1-\textrm{erf}\left(\frac{\alpha}{\eta}\ln\left(\frac{r}{d_{max}}\right)
\right)\right]rdr\nonumber\\
&=&\frac{\pi\rho_L}{2}R^2-\pi\rho_L
\int_{0}^{R}\textrm{erf}\left(\frac{\alpha}{\eta}\ln\left(\frac{r}{d_{max}}\right)
\right)rdr\nonumber\\
&&
\end{eqnarray}
%
%
By employing the substitution $y=\frac{\alpha}{\eta}\ln
\left(\frac{r}{d_{max}}\right)\Rightarrow
r=d_{max}e^{\frac{\eta}{\alpha}y}$, from which
$dr=d_{max}\frac{\eta}{\alpha}e^{\frac{\eta}{\alpha}y}dy$, the
integral (\ref{grado_medio_dv}) takes on the following form:
\[
\int_{0}^{R}\textrm{erf}\left(\frac{\alpha}{\eta}\ln\left(\frac{r}{d_{max}}\right)
\right)rdr=d^2_{max}\frac{\eta}{\alpha}\int_{-\infty}^{I_s}\textrm{erf}(y)e^{2\frac{\eta}{\alpha}y}
dy
\]
whereby,
$I_s=\frac{\alpha}{\eta}\ln\left(\frac{R}{d_{max}}\right)$.

Upon using the following \cite{Gradshteyn}:
\[
\int
e^{ax}\textrm{erf}(bx)dx=\frac{1}{a}\left[e^{ax}\textrm{erf}(bx)-e^{\frac{a^2}{4b^2}}\textrm{erf}\left(bx-\frac{a}{2b}\right)\right],a\ne
0
\]
after some algebra, (\ref{grado_medio_dv}) can be rewritten as
follows:
\begin{eqnarray}\label{grado_medio_dv_4}
E\{d^{NL}_v|R\}&=&\frac{\pi\rho_L}{2}R^2-\pi\rho_L\frac{R^2}{2}\textrm{erf}\left(\frac{\alpha}{\eta}\ln\left(\frac{R}{d_{max}}\right)
\right)\\
&+&\frac{\pi\rho_L}{2}d^2_{max}e^{\frac{\eta^2}{\alpha^2}}\left[1+\textrm{erf}\left(\frac{\alpha}{\eta}\ln\left(\frac{R}{d_{max}}\right)-\frac{\eta}{\alpha}
\right)\right]\nonumber
\end{eqnarray}
Next consider evaluating $E\{d^{NL}_v\}$ over
$\Re^2$. In the limit $R\rightarrow\infty$,
(\ref{grado_medio_dv_4}) simplifies to:
\begin{eqnarray}\label{grado_medio_dv_5}
E\{d^{NL}_v\}=\lim_{R\rightarrow\infty}E\{d^{NL}_v|R\}=\rho_L\pi
d^2_{max}e^{\frac{\eta^2}{\alpha^2}}
\end{eqnarray}
since,
\[
\lim_{x\rightarrow\infty}\textrm{erf}(x)=1
\]
\end{document}